\documentclass[usedcolumn]{mn2e}
\usepackage{times}
\usepackage{epsf}
\usepackage[totalwidth=480pt,totalheight=680pt]{geometry}
\newcommand{\etal}{{et al}\/.}
\newcommand{\PKS}{1610$-$608}
\newcommand{\new}[1]{{#1}}
\newcommand{\newb}[1]{{#1}}
\begin{document}
\title[Jet termination in WATs]{Jet termination in wide-angle tailed
radio sources}
\author[M.J.~Hardcastle and I.~Sakelliou]{M.J.\ Hardcastle$^1$ and I. Sakelliou$^2$\\
$^1$ Department of Physics, University of Bristol, Tyndall Avenue,
Bristol BS8 1TL\\
$^2$ School of Physics and Astronomy, University of Birmingham, Edgbaston, Birmingham B15 2TT
}
\maketitle
\begin{abstract}
Wide-angle tail radio galaxies (WATs) are an uncommon class of radio
sources with luminosities near the FRI/FRII break, and are usually
associated with central cluster galaxies. Their defining
characteristic when imaged sensitively at high resolution is their
twin, well-collimated jets, which can persist with low opening angle
for tens of kpc before flaring into long, often bent plumes. Although
several models for the jet termination have been proposed, the
majority of them are unsatisfactory when confronted with observations.
Here we present the results of a programme of radio observations made
with the aims of showing that objects classified as WATs do all have
well-collimated jets and seeing in detail {\it how} the jets disrupt
as they enter the plumes. We show that compact, `hotspot-like'
features at the ends of the jets are common but by no means universal,
and discuss the constraints that this places on models of the
jet-plume transition. We discuss the properties of the observed
well-collimated jets and, using relativistic beaming models, estimate
their speed to be $\sim 0.3c$. Finally, we show that the distance from
the galactic centre at which the base of the plume is found is related
to the temperature of the host cluster.
\end{abstract}

\section{Introduction}
\label{intro}

The term `wide-angle tail radio galaxy' (WAT) is used in the
literature in a number of different ways. In this paper we use it
(following Leahy 1993) to refer to a clearly identified population of
radio galaxies, which are always associated with a cluster dominant
galaxy, at or near the cluster centre, and are characterized in the
radio by twin, well-collimated jets which may continue for tens of kpc
before abruptly flaring into long, often sharply bent plumes or tails.
\new{The bending of the plumes, the feature which originally gave the
class its name, has long been a source of interest (e.g. Eilek \etal\
1984); a plausible explanation is ram pressure from large-scale bulk
motions in the host cluster due to cluster-cluster mergers (Loken
\etal\ 1995) and this is supported by the facts that the host clusters
of bent sources often show an X-ray structure which is elongated in
the same direction as the WAT tails (G\'omez \etal\ 1997) and that
bent sourcces are often displaced from the X-ray centroids of their
host clusters (Sakelliou \& Merrifield 2000). In this
paper we concentrate on the termination of the well-collimated jets.
We regard this as a separate problem from that of the large-scale
plume bending, both because of the existence of relatively straight
objects that are WATs by our definition (e.g. 3C\,130, Hardcastle
1998) and because there are well-known cluster-centre objects that do
not contain WAT-type jets but show very similar large-scale bending
(e.g. 3C\,75, Eilek \& Owen 2002; Appendix A).}

\new{The termination of the WAT jets is an outstanding problem with} our
present understanding of how radio sources interact with the
intracluster medium. In the (few) well-studied objects the jets
resemble those in classical double (FRII) radio sources, in which the
jet termination is due to a strong shock as jet plasma interacts with
the external medium, giving rise to a hotspot. But in WATs, which have
luminosities similar to those of FRIIs, the jets terminate not at the
very ends of the sources, but near the base of the plumes. The
outstanding question is therefore: what causes the jets to disrupt?
\new{This is also a key question if we wish to test models that try to
explain the bending and distortion of the plumes: it is the physical
processes that take place at the jet termination that define the
plumes' nature and characteristics.}

Because WATs are rare compared to other classes of radio galaxies,
there have been few detailed radio observations which might help to
\new{explain why WAT jets disrupt}. Radio observations of samples of
WATs (e.g.\ O'Donoghue, Owen \& Eilek 1990) have concentrated on their
large-scale structure and the dynamics of plume bending. Images of
this kind usually show a well-collimated inner jet, and typically show
a bright `flare point' at the base of the plume, sometimes referred to
(e.g.\ O'Donoghue, Eilek \& Owen 1993) as a hotspot, but do not have
enough resolution to resolve it transversely and give structural
information on the jet termination; thus, in general, we cannot say
{\it how} the jets disrupt. However, Hardcastle (1998) showed that the
WAT 3C\,130 ($z=0.109$) exhibits a bright, compact, sub-kpc feature at
the termination of the N jet. The similarity of this feature to the
terminal hotspots in FRIIs led Hardcastle (1998) to suggest that
3C\,130's N jet also terminated in a shock; further multi-frequency
observations suggest that the absence of a comparable feature in the S
plume is reflected in the source's particle acceleration properties
(Hardcastle 1999). These observations motivated us to carry out a
comparably detailed study of a larger sample of WATs, with the aim of
determining the properties of the inner jets and the nature of the jet
termination in the WAT population as a whole. In this paper we present
the new radio observations and discuss some conclusions that can be
drawn from them.

Throughout the paper we use a cosmology with $H_0 = 65$ km s$^{-1}$
Mpc$^{-1}$, $\Omega_{\rm m} = 0.3$ and $\Omega_\Lambda = 0.7$.
Spectral indices $\alpha$ are defined
in the sense $S \propto \nu^{-\alpha}$. B1950.0
co-ordinates are used.

\section{Sample, observations and analysis}
\label{radio}

The sample we chose to observe in the radio was a subset of the sample
of Sakelliou \& Merrifield (2000), who based their object list on the
radio survey of Abell clusters by Owen \& Ledlow (1997, and references
therein). From that sample we chose a few objects that were
approximately at the centres of their host clusters and that clearly
showed the flaring at the base of the plumes that characterizes the
class, although not all of them were known to have faint, narrow inner
jets. There was otherwise no morphological selection. To supplement
this sample we added one southern source, \PKS , which was not in Owen
\& Ledlow's survey because of its declination but which otherwise met
our selection criteria. The properties of the observed sample are
listed in Table \ref{observed}. \newb{We emphasise that this is in no
sense a statistically complete sample, and that it is possible that
our morphological selection, based on low-resolution maps, has
introduced some bias that renders these objects
unrespresentative of the parent population of WATs as we define them.
However, we hope that the structures we observe in jet terminations in
the sample should be representative of those of the WAT population as
a whole. As we will discuss later (Section 4.1) our observations give
us reason to believe that WATs (according to our definition) can indeed be
safely identified from low-resolution radio maps.
}

\begin{table*}
\caption{The observed sample}
\label{observed}
\begin{tabular}{llr....r}
\hline
Source name&Other name&Abell&z&\multicolumn{1}{c}{Angular
scale}&\multicolumn{1}{c}{1.4-GHz flux}&\multicolumn{1}{c}{$\alpha_{1.4}$}&\multicolumn{1}{c}{1.4-GHz luminosity}\\
&&cluster&&\multicolumn{1}{c}{(kpc arcsec$^{-1}$)}&\multicolumn{1}{c}{density (Jy)}&&\multicolumn{1}{c}{(W Hz$^{-1}$ sr$^{-1}$)}\\
\hline
0647+693&4C\,69.08&562&0.11&2.16&0.800&1.03&$2.3 \times 10^{24}$\\
1231+674&4C\,67.21&1559&0.1049&2.07&0.879&0.88&$2.3 \times 10^{24}$\\
1333+412&4C\,41.26&1763&0.2278&3.93&0.797&1.08&$1.2 \times 10^{25}$\\
1433+553&4C\,55.29&1940&0.1402&2.66&0.447&0.79&$2.1 \times 10^{24}$\\
2236$-$176&PKS&2462&0.0742&1.52&1.642&0.78&$2.0 \times 10^{24}$\\
3C\,465&&2634&0.0302&0.65&7.650&0.82&$1.5 \times 10^{24}$\\
\PKS&PKS&3627&0.0157&0.34&48&1.24&$2.5 \times 10^{24}$\\
\hline
\end{tabular}
\begin{minipage}{14cm}
\vskip 5pt Redshifts are taken from Owen, Ledlow \& Keel (1995),
except for that of 0647+693 (Stickel \& K\"uhr 1994) and \PKS\ (the
cluster redshift, corrected for Local Group motions, from Struble \&
Rood 1999). 1.4-GHz flux densities are from Ledlow \& Owen (1995),
except for \PKS, where the flux density is from Christiansen
\etal\ (1977). Spectral indices, are determined from lower-frequency
observations where possible: 6C 151-MHz flux densities for 0647+693,
1231+674, 1333+412 and 1433+553, the MRC 408-MHz flux density for
2236$-$176, the 3CRR 178-MHz flux density for 3C\,465, and the 843-GHz
MOST observations of Jones \& McAdam (1992) for \PKS. These spectral
indices are used to calculate the 1.4-GHz luminosity in the source
frame.
\end{minipage}
\end{table*}

We chose to observe the northern sources in our sample using the NRAO
Very Large Array (VLA) at 8.5 GHz. This observing frequency was chosen
because of the high angular resolution it allows (up to $\sim 0.25$
arcsec in the A configuration of the VLA) and because the use of high
frequencies mitigates the potentially important effects on the
sources' polarization properties of Faraday rotation due to the
intra-cluster medium (discussed further below). 

Since the main aim of our observations was high angular resolution,
the majority of the newly allocated observing time was spent in the A-
and B-configurations of the VLA, with only short observations at C-
and D-configurations. All our new VLA observations used two observing
frequencies at 8.44 and 8.49 GHz, with a bandwidth of 25 MHz (for the
A-configuration observations, to avoid bandwidth smearing) and 50 MHz
(for the rest). However, we obtained C- and D-configuration data for
half of our sources from the VLA archive, making use of long existing
observations at similar frequencies taken for other purposes and
published elsewhere (Katz-Stone \etal\ 1999; Eilek \& Owen 2002).
Details of the VLA observations used are given in Table \ref{vlaobs}.
The initial data reduction was carried out in {\sc aips} in the
standard manner. Individual datasets were phase self-calibrated to
convergence, where enough {\sc clean} components were available. In
3C\,465 and 1433+553 there was some evidence of core variability over
the epochs of observation, and so the task {\sc uvsub} was used to
bring the core flux densities into agreement. The datasets were then
cross-calibrated and merged, beginning with the highest-resolution
data (so that B-configuration data were cross-calibrated with
A-configuration data and merged, C-configuration data cross-calibrated
with the resulting A+B dataset and merged, and so on) to give a final
dataset containing phase-aligned A,B,C and D-configuration data. We weighted
down the short-baseline datasets in the cases with long C- and
D-configuration observations in order to avoid being dominated by
these data in the imaging stages; otherwise reweighting on merging
reflected only the bandwidth of the observations.

\begin{table*}
\caption{Observations made with the VLA}
\label{vlaobs}
\begin{tabular}{llrlrlrlr}
\hline
Source&\multicolumn{2}{c}{A configuration}&\multicolumn{2}{c}{B
configuration}&\multicolumn{2}{c}{C
configuration}&\multicolumn{2}{c}{D configuration}\\
&Date&Time (h)&Date&Time (h)&Date&Time (h)&Date&Time (h)\\
\hline
0647+693&2001 Jan 15&2.0&2001 May 19&1.5&2001 Jul 19, 29&1.0&2001 Dec 19, 20&0.5\\
1231+674&2001 Jan 15&2.0&2001 May 19&1.5&1992 Apr 30$^a$&3.0&1992 Aug 6, 11, 18$^a$&4.6\\
1333+412&2001 Jan 16&2.0&2001 May 19&1.5&2001 Jul 29&1.0&2001 Dec 20&0.5\\
1433+553&2001 Jan 16&2.0&2001 May 19&1.5&1992 Apr 30$^a$&3.0&1992 Aug 6, 11, 18$^a$&4.3\\
2236$-$176&2001 Jan 15&2.0&2001 May 19&1.5&2001 Jul 19&1.0&2001 Dec 19&0.5\\
3C\,465&2001 Jan 15&2.0&2001 May 19&1.5&1989 Sep 20$^b$&6.7&1989 Dec 01$^b$&1.3\\
\hline
\end{tabular}
\begin{minipage}{15.7cm}
Notes: $^a$ Observations made by D.M.\ Katz-Stone at an observing
frequency of 8.41 GHz. $^b$ Observations made by J.A.\ Eilek at observing
frequencies of 8.0 and 8.5 GHz. All observations, except for those at
A-array, used two channels with a bandwidth of 50 MHz.
\end{minipage}
\end{table*}

The southern source \PKS\ was observed with the ATCA in two
configurations, as shown in Table \ref{atca}, at 4.80 and 8.64 GHz.
The initial data reduction for this source was carried out in {\sc
miriad}, and again followed standard procedures. The data were then
self-calibrated, cross-calibrated, and merged in {\sc aips}. Strong
RFI during the second observing run necessitated significant data
flagging, so the overall $u,v$ coverage is not as good as might be
expected.

The final imaging was carried out in {\sc aips} using the task {\sc
imagr}, combining the two observing frequencies (in the case of the
VLA data) or the 13 spectral channels (in the case of the ATCA data)
to make single images, which for the VLA data are at an intermediate
effective frequency of 8.47 GHz. The $u,v$ plane was weighted (using
the ROBUST and UVTAPER parameters of {\sc imagr}) to obtain the
desired resolution. Images, in all cases based on the full combined
datasets, are presented in the following section.

\begin{table}
\caption{ATCA observations of \PKS}
\label{atca}
\begin{center}
\begin{tabular}{llr}
\hline
Configuration&Date&Duration (hours)\\
\hline
6A&2000 Oct 07&9.6\\
1.5E&2000 Nov 24&7.8\\
\hline
\end{tabular}
\end{center}
\end{table}

\section{Images}

In this section we present total-intensity and polarization images of
each of the sources in our sample. In most cases, a low-resolution
image is presented to show the large-scale structure of the source;
high-resolution images of the jet terminations can then be seen in
context. Properties of the maps presented are given in Table
\ref{maps}. We comment briefly on each source in turn.

A note of caution is necessary regarding polarization maps. At our
observing frequency, Faraday rotation is negligible (involving
corrections to the polarization position angle of $\la5^\circ$) if the
magnitude of the rotation measure (RM) towards a particular point on
the source is $\la 70$ rad m$^{-2}$. If Faraday rotation is
negligible, then the direction of the $B$-field in the incoming radio
waves, which is the direction of the vectors we plot in all the
polarization maps that follow, tells us the (emission-weighted mean)
magnetic field direction in the synchrotron-emitting plasma. However,
all our sources lie in cluster environments, and several of them are
in rich clusters for which RM magnitudes higher than $70$ rad m$^{-2}$
are frequently observed. The only direct observational evidence we
have on individual sources (except for \PKS, which we discuss in more
detail below) comes from the observations of 3C\,465 by Eilek \& Owen
(2002), which show the magnitude of RM to range \new{up to $\sim 250$ rad
m$^{-2}$ in small, discrete regions}: this would imply polarization corrections of around
$20^\circ$. However, 3C\,465's host cluster
may not be representative. It should therefore be borne in mind that
the polarization vectors we plot are only the best estimate available
of the true $B$-field directions, although for simplicity we describe
them as magnetic field vectors in what follows.

The reader is reminded that we use the term `jets' only for the inner,
well-collimated structures in the WATs. The brighter, broader features
which dominate the sources' radio emission are referred to as
`plumes'.  The central unresolved feature, coincident with the nucleus
of the host galaxy, is the `core' and we tabulate core properties,
measured from high-resolution maps, in
Table \ref{cores}.

\begin{table*}
\caption{Properties of the maps presented}
\label{maps}
\begin{tabular}{lrrrrrr}
\hline
Source&\multicolumn{3}{c}{Restoring
beam}&\multicolumn{2}{c}{Off-source noise}&Figure\\
&Major axis&Minor axis&Pos. angle&Stokes $I$&Stokes $Q$, $U$&number\\
&(arcsec)&(arcsec)&(degree)&($\mu$Jy)&($\mu$Jy)\\
\hline
0647+693&2.03&1.84&73&24&25&\ref{0647l}\\
&0.90&0.61&59&23&&\ref{0647h}\\[2pt]
1231+674&2.38&1.98&$-$62&11&11&\ref{1231l}\\
&1.24&0.73&$-$58&12&12&\ref{1231h}\\[2pt]
1333+412&1.28&0.81&$-$65&15&15&\ref{1333l}\\
&0.38&0.27&$-$76&13&&\ref{1333h}\\[2pt]
1433+553&2.32&1.81&$-$59&12&12&\ref{1433l}\\
&1.04&0.55&$-$65&12&13&\ref{1433h}\\[2pt]
2236$-$176&2.59&1.82&10&18&15&\ref{2236l}\\
&0.78&0.52&23&13&15&\ref{2236h}\\[2pt]
3C\,465&2.60&2.03&$-$68&26&12&\ref{465l}\\
&0.61&0.60&53&26&20&\ref{465h}\\[2pt]
\PKS&2.52&1.78&$-$10&740&300&\ref{1610}\\
&1.30&1.01&3&430&78&\ref{1610}\\
\hline
\end{tabular}
\end{table*}

\begin{table}
\caption{Properties of the radio cores of the sample}
\label{cores}
\begin{center}
\begin{tabular}{l.rr}
\hline
Source&\multicolumn{1}{c}{Core flux}&RA (B1950)&DEC (B1950)\\
&\multicolumn{1}{c}{density (mJy)}&\\
\hline
0647+693&4.9&06 47 54.72&+69 23 31.2\\
1231+674&5.5&12 31 02.58&+67 24 15.6\\
1333+412&3.3&13 33 09.64&+41 15 22.6\\
1433+553&15.1^*&14 33 54.83&+55 20 54.2\\
2236$-$176&11.8&22 36 30.22&$-$17 36 06.8\\
3C\,465&214^*&23 26 00.44&26 45 21.26\\
\PKS&130&16 10 43.08&$-$60 46 54.2\\
\hline
\end{tabular}
\vskip 5pt
\begin{minipage}{7.3 cm}
Core fluxes quoted are the integrated values from Gaussian fitting to
high-resolution maps using the {\sc aips} task {\sc jmfit}; the
positions quoted are also {\sc jmfit} output. Flux densities marked
with an asterisk indicate that the core had detected variability
between the epochs of our observations.
\end{minipage}
\end{center}
\end{table}

\subsection{0647+693}

No detailed radio observations of this source had previously been
made. Based on a snapshot image, O'Dea \& Owen (1985) remarked that
the source's structure is intermediate between narrow-angle and
wide-angle tails, an impression reinforced by our data. G\'omez \etal\
(1997) showed that the radio source is significantly (30 arcsec) to
the west of the cluster X-ray peak. DSS2 images show that the host
galaxy is also 21 arcsec to the west of a second bright elliptical,
although the host does appear to be the brighter of the two by around
0.5 mag in $R$.

Our low-resolution maps (Fig.\ \ref{0647l}) show that the plumes,
particularly the southern plume, are not abnormal for a WAT. However,
the high-resolution maps (Fig.\ \ref{0647h}) show that the inner jets
are very much more bent than in any of the other objects in our
sample; the southern jet is detected over almost all its length, and
(in projection) makes two sharp, almost $90^\circ$ bends at the
positions labelled SJ1 and SJ2 in Fig.\ \ref{0647h}. The northern jet
is less well detected but appears to have bent in a similar way by the
time it becomes visible at NJ1. If the jet is ballistic in this source
(cf.\ Blandford \& Icke 1978), the strong bending would require the
source to have been moving eastwards on average almost as fast as the
jet flow (and faster at some times) which seems unlikely (we discuss
jet speeds in more detail below, section \ref{speeds}). Instead, it
is tempting to attribute both the jet bending and the general
direction of the plumes to ram pressure as a result of bulk motion of
the host galaxy towards the cluster centre and/or the neighbouring
elliptical.

The N jet cannot be reliably traced into the N plume, but it seems
most likely that it terminates in the compact (arcsec-scale,
comparable in size to the jet) feature
N1 at the N edge of the N plume. In the S plume it is not clear
whether the jet terminates in S1 or S2, but S2 contains the most
convincing compact structure.

\subsection{1231+674}

This source has previously been imaged by O'Donoghue \etal\ (1990) and
Katz-Stone \etal\ (1999). The image of the large-scale structure
(Fig.\ \ref{1231l}) thus shows few new details. The arc appearing
perpendicular to the S plume, commented on by O'Donoghue \etal , turns
out to have an unresolved radio core identified with an optical galaxy in
the field visible in the DSS2 plates. It is thus most likely to be a
narrow-angle tail (NAT) radio galaxy falling into the potential well of
A1559. The host galaxy of 1231+674 itself is a dumb-bell system
aligned approximately E-W on the sky, with the radio source being
associated with the brighter western component.

The high-resolution image (Fig. \ref{1231h}) shows two well-defined
jets. The N jet bends smoothly and terminates in the bright region N1;
this region contains some compact structure, elongated in the jet direction,
whose magnetic field direction is transverse to the jet direction. The
S jet bends abruptly at two knots, S1 and S2, in which the magnetic
field is transverse to the jet direction, but then there appears to be
continued well-collimated outflow, with a longitudinal magnetic field,
from S2 to the eventual jet termination in the diffuse region S3.
1231+674, like 3C\,130, is a source in which the jet termination is
not associated with significant observed bending in the plumes.

\subsection{1333+412}

This small source has not been observed previously at high resolution;
the best previous maps are those of Owen \& Ledlow (1997). Our
low-resolution image (Fig.\ \ref{1333l}) shows much the same features,
but gives a clear detection of the inner jets. The small source to the
E is associated with one of the many fainter galaxies in the field,
and is likely to be a NAT lying in the cluster. The N lobe is
noticeably less strongly polarized than the S lobe. Since the cluster
environment of 1333+412 is known to be rich, this is plausibly due to
depolarization from the intra-cluster medium (ICM).

The high-resolution image (Fig.\ \ref{1333h}) shows the S jet
penetrating into the plume for some distance (SJ1) before it terminates
in the compact region S1. The current termination position of the N
jet is not clear; it points towards either N2 or N3, but the brightest
feature in the plume is N1.

\subsection{1433+553}

This source was previously observed by O'Donoghue \etal\ (1990) and
Katz-Stone \etal\ (1999). The image of the large-scale structure
(Fig.\ \ref{1433l}) is similar to others previously presented. The S
lobe is unusual in that it shows no obvious evidence, either in total
intensity or polarization, for a plume leading away from the source.
It is possible that the plume makes a small angle to the line of
sight, so that we are seeing it in projection.

The high-resolution image (Fig.\ \ref{1433h}) shows the S jet bending
(SJ1) before it enters a reasonably compact bright region, S1. The N
jet is not traced all the way into the N plume at this resolution, but
there is a clear compact feature (N1) in the plume with an unusual
`shock-like' structure.

\subsection{2236$-$176}

This source was previously observed by O'Donoghue \etal\ (1990), and
our low-resolution image (Fig.\ \ref{2236l}) shows similar features;
the noteworthy feature of the source is the $90^\circ$ bend in the E
plume. On smaller scales, our high-resolution image (Fig.\
\ref{2236h}) shows a well-collimated, asymmetrical pair of jets (NJ1,
SJ1) which bend with S-symmetry at about the same distance from the
nucleus (NJ2, SJ2). The fainter N jet appears to retain its
collimation (NJ3) until it disappears into a diffuse region of
high-surface-brightness emission (N1) with no obvious compact features
at the base of the plume. The S jet is similar, although there is a
faint compact feature (S1) which may be associated with the jet
termination.

\subsection{3C\,465}

3C\,465 has been imaged in several earlier studies, including Leahy
(1984), Eilek \etal\ (1984) and, most recently, Eilek \& Owen (2002).
Its large-scale properties (Fig.\ \ref{465l}) are well known.

At high resolutions (Fig.\ \ref{465h}), knotty structure in its
one-sided jet is visible (NJ1) but there is no compact structure after
the jet bends W at NJ2, although a distinct region downstream (N1) is
visible in total intensity and polarization. The
high-surface-brightness base of the S plume is well resolved, showing
filamentary structures (S1, S2) that are not obviously related to the
termination of the S jet. 

\subsection{\PKS}

\PKS\ lies in the nearby massive cluster Abell 3627 (B\"ohringer \etal\
1996) which has been suggested as the core of the so-called `Great
Attractor' (e.g.\ Kraan-Korteweg \etal\ 1996). The radio source was shown to
have WAT-like structure by the low-resolution ATCA observations of
Jones \& McAdam (1996); its proximity gives it a claim to be called
the closest WAT source. Ours are the first high-resolution
observations. Because we do not have short-baseline data, the
large-scale structure of the source is not shown. However, our
observations (Fig. \ref{1610}) clearly show the
characteristic inner jets of the WAT class. Neither jet appears to
terminate in any particularly compact structure. The inner regions of
the source are quite strongly polarized, and there is little evidence
for depolarization between the two frequencies, but the position angles of
the polarization vectors differ markedly between 5 and 8 GHz,
particularly in the E plume, suggesting that a resolved Faraday screen
is present.

\section{Radio properties of wide-angle tails}

\subsection{WAT morphology}

All the sources in our sample clearly show the narrow, well-collimated
inner jets, leading from the nucleus to the broader plumes, that we
regard as a defining characteristic of the class; in some cases
(0647+693, 1333+412, 1610$-$608) the jets are detected for the first
time by our observations. This illustrates that WATs can reliably be
identified from low-resolution images on the basis of their `flaring'
morphology. \newb{We find that many of the jet-plume transitions are not
associated with significant plume bending; in fact, if we exclude
0649+693, 9/12 of the plumes in the present sample do not apparently bend
strongly at or near the transition. It seems unlikely that such a
large number of relatively straight jet-plume transitions could be
explained by fortuitous projection effects.}

The main surprise from the point of view of source structure was the
inner jets in 0647+693, with their clear large-scale bending. We
suspect that there has been a bias against such objects in previous
high-resolution radio imaging work, since a number of candidate
objects with similar plume behaviour can be identified in the images
of Owen \& Ledlow (1997); we hope to investigate their inner jet
properties with future VLA observations. The existence of the hotspot,
and the well-collimated jets in this source, \new{provide an
additional argument} against models that invoke ram pressure to
explain the jet termination or jet disruption\new{; these jets appear
to be bending due to ram pressure well before their termination, which
would not be expected in the models of Loken \etal\ (1995). This,
\newb{together with the observation that plumes are often initially
straight}, reinforces earlier arguments (Section \ref{intro}) that the
plume bending and jet termination in WATs are separate and unrelated
problems.}

It is worth noting that the magnetic field direction in the jets is
generally parallel to the jet direction, as we might expect for
FRII-like jets. \new{In more normal FRI jets, there is typically a
transition from a parallel to a perpendicular field configuration,
which usually occurs on scales of a few kpc (see Appendix A for an
example).}

\subsection{Speeds and beaming}

\label{speeds}
Our good images of the jets allow us to determine jet/counterjet
ratios (jet sidednesses) for the sample. These are listed in Table
\ref{sided}; we follow the procedure defined by Hardcastle \etal\
(1998) and measure only the straight part of the jet, to ensure that
there is no change in the angle it makes with the line of sight over
the integration region. As O'Donoghue \etal\ (1993) remarked (with a
sample containing several sources in common with ours) the jet
sidednesses are generally close to unity. If we assume that the jets
are intrinsically symmetrical and that the jet sidedness ratio is due
to relativistic beaming effects, we can constrain the characteristic
beaming speed $\beta$. There is no direct independent evidence for
relativistic speeds in these jets, although it is interesting that the
two sources that show variable radio cores (an indication of beaming)
also show comparatively one-sided jets. Nevertheless, it seems
plausible that beaming accounts for a substantial fraction of the
jet-counterjet asymmetry. The maximum-likelihood value of $\beta$, for
the jets (if we assume that they sample all possible angles to the
line of sight and that the jets are intrinsically symmetrical) is
around 0.3; this is also true for the O'Donoghue \etal\ objects and
for the combination of the two samples, for which we find $\beta = 0.3
\pm 0.1$, determining errors using a Monte Carlo technique. These WAT
characteristic speeds are thus significantly lower than those measured
by Wardle \& Aaron (1997) for FRII quasars, or Hardcastle \etal\
(1999) for FRII radio galaxies, but it is not clear that the WAT
sample is unbiased. It would be desirable to repeat this speed
determination for a larger sample, selected purely on WAT morphology
and with homogeneous jet sidedness measurements.

If the characteristic jet speeds are in fact lower than those of
FRIIs, it may indicate a lower bulk speed throughout the jet, but
would seem more likely to be a result of a stronger interaction
between the jet and its environment in the case of WAT jets which are
not protected by lobes, leading to a brighter, slower
shear layer at the boundary of the jet. We note that some FRII jets
which appear to lie outside their lobes are similarly bright and
two-sided (e.g. 3C\,438, Hardcastle \etal\ 1997). If WAT jets are
relativistic in general, then ballistic models for the jet wiggles
seen in many sources, and for the bends seen in 0647+693, are untenable;
instead, these must be evidence for interaction between the jet and
features of the external environment (either due to thermal or ram
pressure). This in turn requires that the jets be light compared to
the external medium, since the external sound speed is of the order of
1 per cent of the beaming speed (which, if it is due to a dissipative
shear layer, may itself be an underestimate of the mean bulk speed of
the flow).

\begin{table}
\caption{Straight jet flux densities and sidedness ratios}
\label{sided}
\begin{center}
\begin{tabular}{l..r}
\hline
Source&\multicolumn{1}{c}{Jet flux}&\multicolumn{1}{c}{Counterjet}&Sidedness\\
&\multicolumn{1}{c}{(mJy)}&\multicolumn{1}{c}{flux (mJy)}&ratio\\
\hline
0647+693 & 0.60 & 0.13 & 4.79$^*$\\
1231+674 & 2.26 & 1.84 & 1.23\\
1333+412 & 0.93 & 0.81 & 1.15\\
1433+553 & 1.17 & 0.33 & 3.55\\
2236$-$176 & 17.94 & 8.65 & 2.08\\
3C\,465 & 22.46 & 4.21 & 5.34\\
1610$-$608 & 48.64 & 40.60 & 1.20\\
\hline
\end{tabular}
\vskip 5pt
\begin{minipage}{6cm}
$^*$ Note that the straight jet region in this source is very short, so
that the sidedness ratio may not be meaningful.
\end{minipage}
\end{center}
\end{table}

\subsection{Jet termination}

Our observations allow us to give a qualitative answer to the question
posed in Section 1; how do the jets terminate? In fact, the larger
sample gives results consistent with those already seen in 3C\,130:
jets in WATs terminate in a variety of ways, and do not necessarily
show a compact `hotspot'. There are compact (comparable in size to the
jet width), relatively bright features associated with the jet
termination in 0647+693 (N jet), 1231+674 (N jet), 1333+412 (both
jets) and 1433+553 (possibly both jets) but not clearly in 2236$-$176,
3C\,465 or 1610$-$608, in which the jets simply make a smooth
transition into the base of the plumes. Hotspots, where present, are
not typically at the base of the plume, but some distance in,
suggesting that the jet continues to be well collimated for some time
after it enters the plume. Several sources, of which 1231+674 S is the
best example, have jets that apparently penetrate undisrupted into the
plumes for some distance before disappearing. This behaviour, also
seen in 3C\,130 S (Hardcastle 1998) seems to rule out models in which
the jet disruption is related to an externally produced shock, or to
the propagation of the jet across some other structure in the ICM; in
such models the jet termination position would be fixed, so that there
would be no reason for the plume to extend closer to the core than the
end of the jet.

All the observations are, however, consistent with the idea
(Hardcastle 1999) that the jets are `flapping' in the bases of the
plumes and show compact jet termination features only when they happen
to impinge on the plume's edge. The northern `hotspot' of 0647+693 is
particularly suggestive in this respect, as it lies on the outer edge
of the plume. In this picture, the jet terminates where it does {\it
now} because it enters the plume base and encounters the plume edge.
If the structure of WATs is due to the properties of the ICM, then
these must determine not the current jet termination point (which is
ephemeral in any case) but the location of the base of the plume.

\section{Jet termination and cluster properties}

We were struck in carrying out this analysis by the wide range in the
(projected) lengths of the inner jets that our target sources exhibit.
If their cluster environments are responsible for the unique structure
of WATs, then radio source properties such as jet length or distance
to the plume base should be related to cluster richness, which we can
parametrize in terms of \new{the overall `cluster temperature'
  determined from low-resolution X-ray observations. We emphasise
  that we are simply trying to parametrize cluster richness by using
  this quantity; in general clusters will have temperature gradients
  and the central regions of the cluster may well be cooler than the
  value we quote.} The advantage of
working with a sample of Abell clusters is that X-ray observations
already exist for many of our sources. Accordingly, we collated
cluster temperatures from the literature for our targets; X-ray
measurements were available for all but one (1231+674). To supplement
our small sample, we also investigated the other sources from the
Sakelliou \& Merrifield (2000) sample that would have met our
morphological selection criteria, listing those for which temperature
measurements were available, and we also include the well-known WAT
sources 3C\,130 (Hardcastle 1998), Hydra A (Taylor \etal\ 1990),
4C34.16 (Sakelliou, Merrifield \& McHardy 1996), Abell 2029 (Taylor, Barton \&
Ge 1994) and Abell 160 (Jetha \etal\ in prep.). We define the jet
termination length as the mean of the two linear distances between the
core and the compact `hotspot' in the base of the plume, or the
brightest feature in the base of the plume if no compact feature
exists. The temperatures and termination lengths of the objects in the
combined `sample' (which is in no sense complete, but should not be
biased in any obvious way) are listed in Table \ref{ctemp}.

\begin{table*}
\caption{WAT jet termination lengths and cluster temperatures for the
  sample and for WAT sources from the literature}
\label{ctemp}
\begin{tabular}{ll.rr.r}
\hline
Radio name&Abell cluster&$z$&Jet termination&Reference&\multicolumn{1}{r}{Cluster temp}&Reference\\
&&&length (kpc)&for radio&\multicolumn{1}{r}{(keV)}&for $kT$\\
\hline
0647+693&562&0.11&81&1&1.7&2\\
1231+674&1559&0.1049&35&1&-\\
1333+412&1763&0.2278&20&1&6.9&3\\
1433+674&1940&0.1402&49&1&1.6&2\\
2236$-$176&2462&0.0742&44&1&1.5&2\\
3C\,465&2634&0.0302&28&1&3.5&4\\
\PKS&3627&0.0157&13&1&7&5\\
\hline
3C\,130&--&0.109&58&6&2.9&6\\
Hydra A&--&0.052&11&7&4.0&8\\
4C34.16&--&0.078&51&9&3.2&9\\
0110+152&160&0.0445&42&9&2.2&9\\
3C\,40&194&0.018&59&10&2.6&11\\
1159+583&1446&0.104&19&12&2.6&2\\
1508+059&2029&0.076&3&12&7.8&3\\
1636+379&2214&0.1610&35&13&2.3&2\\
\hline
\end{tabular}
\vskip 5pt
\begin{minipage}{12cm}
The horizontal line divides sources imaged in this paper from sources
whose properties are taken from the literature. References are (1) This paper (2) G\'omez \etal\ 1997 (3) David \etal\ 1993 (4)
Schindler \& Prieto 1997 (5) Tamura \etal\ 1998 (6) Hardcastle 1998
(7) Taylor \etal\ 1990 (8) David \etal\ 1990 (9) Unpublished {\it
  Chandra}, {\it XMM} and VLA data (10) O'Dea \& Owen 1985 (11) Fukazawa \etal\
1998 (12) Taylor \etal\ 1994 (13) O'Donoghue \etal\ 1990
\end{minipage}
\end{table*}

Cluster temperature is plotted as a function of jet termination length
in Fig.\ \ref{ktlen}. There is a clear trend in this plot, in the
sense that hotter clusters have systematically shorter jets. The
inverse correlation is significant at the 99 per cent confidence level
on a two-tailed Spearman rank test. Although there is significant
scatter in the correlation -- unsurprisingly, given that all the
distances should be corrected for an unknown projection factor and
that the errors, not plotted for clarity, on the X-ray temperatures
are sometimes large -- its significance leads us to believe that we
are seeing a real physical effect. Although it has been suggested that
WATs are rare in `cooling flow' clusters, two of the sample in Table
\ref{ctemp} (Hydra A and A2029) would traditionally have fallen into
this category\footnote{A2029 has been shown by Lewis, Stocke \& Buote (2002)
to have no spectral evidence for cooling X-ray emitting gas, and by
Baker \etal\ (2003) to have no H$\alpha$ recombination emission, and
so is unlikely to have a high deposition rate of cold matter.}. We find
it interesting that the WATs in these clusters, though both small,
seem to lie close to the relation established for non-`cooling flow'
systems.

We consider the consequences of the length-temperature correlation for
two WAT formation models.

(1) {\it Jet disruption at an interstellar medium/ICM interface}
    (Sakelliou \& Merrifield 1999). In this model, the plumes are
    generated when the jet crosses the ISM/ICM interface. The steep
    density change in their environment possibly encourages the growth
    of large-scale instabilities in the shear layer of the jet; as our
    observations show, in some cases the jet does not disrupt at the
    base of the plume, but appears to carry on inside the plume.
    Detailed analysis of the properties of the environment of the jets
    and plumes will be able to shed more light on this issue. If the
    above is true, the observed distance between the core and the
    plume base gives a measurement of the size of the galactic
    atmosphere. The correlation shown in Fig.\ \ref{ktlen} would imply
    that the ISM of the central cluster galaxy is smaller in richer
    clusters. At first sight, this result seems to contradict common
    notions about the hot ISM in elliptical galaxies. Intuitively, one
    might expect the cores of richer clusters to be occupied by larger
    galaxies (as a result of higher galaxy merging rate) and so would
    think that they should have larger ISMs. Earlier studies have
    indeed found positive correlations between the optical luminosity
    and temperature (and X-ray luminosity) for elliptical galaxies
    (e.g.\ Edge \& Stewart 1991). However, these studies have largely
    been restricted to either isolated elliptical galaxies (or
    galaxies in poor environments), or X-ray faint galaxies, whose
    X-ray emission should be dominated by the stellar population. It
    is not clear that one should assume that the ISM in BCGs should
    also follow the stellar distribution. A plausible argument is that
    the correlation in Fig.\ \ref{ktlen} is caused by the higher
    pressures in the centres of rich clusters, which would naturally
    cause a hot ISM of a given mass to have a smaller radius.
    \new{Another possibility is that the higher ram pressure from
    large-scale bulk motions in richer clusters is more efficient at
    stripping the ISM from the central cluster galaxy.} We therefore
    cannot rule out the idea that the jet disrupts at the ISM/ICM
    interface, based on this correlation, \new{although it does not
    give us any new information on {\it why} the jet should disrupt at
    this location}. However, we would then
    expect a relationship between the temperature and the interface
    radius to hold in all BCGs, not just WAT hosts, which is a
    testable prediction of the model.

(2) {\it WATs as failed FRIIs}. In the `cocoon-crushing' model
    proposed by Hardcastle (1999), WATs are formed when the lobes of
    FRIIs are driven out to large radii by buoyant forces as the
    internal lobe pressure drops, exposing the jet to the external
    medium. This would happen only for weak FRIIs, in which the
    expansion speed, driven by the internal lobe pressure, became
    subsonic (i.e. the internal and external pressures became
    comparable) while the source was still close to the cluster
    centre. The speed at which buoyant forces can move radio-emitting
    plasma is determined by the external sound speed and the central
    density gradient: under free acceleration, the time taken to move
    a component of the source a given distance would be inversely
    proportional to temperature, with a relation of the form
\[
t \propto \left({{kT \nabla \rho}\over{\rho}}\right)^{-1/2}
\]
Both the external pressure and the sound speed
    are higher in rich, hot clusters, and so for a given jet
    power and a fixed density profile we would expect this process to
    take place earlier in the parent FRII's life, making the resulting
    WAT shorter. Qualitatively, therefore, the observation that the
    plume bases are closer to the nucleus in hotter clusters is
    consistent with the failed-FRII model, although it does require
    that the jet luminosities of all the sources should be broadly
    similar. We note that this model would also predict stronger
    buoyant forces, and therefore a shorter core-plume base distance,
    in sources where the central density gradient was particularly
    high: this might explain why Hydra A, which lies in a well-known
    `cooling flow' cluster, has a very short core-plume base distance
    even though the cluster temperature is not very high (Table \ref{ctemp}).

\section{Conclusions}

We have detected well-collimated, bright
inner jets in all of our small sample of WATs. The termination of
these inner jets is often, but not universally, associated with a compact,
hotspot-like radio feature. By analogy with the similar features in
FRII radio galaxies, we may associate these compact features with
shocks at which the jet interacts with some stationary medium. The
fact that some sources {\it do not} show such termination features,
but instead have jets that propagate undisrupted for some distance
into the plumes, strongly suggests that the {\it current} location of the
jet termination is not associated with some feature of the external
medium, such as propagation across a shock front (Norman \etal\ 1988)
or the ISM/ICM interface (Loken \etal\ 1995). Instead, it seems likely
that the jet terminates where it does as a result of interaction with
small-scale bulk motions of the plasma in the cluster and inside the
plume base, which causes the end of the jet to move around inside the
plume. Jet termination shocks are plausibly observed only when the jet
happens to intersect with the edge of the plume. Jets which do not
undergo jet termination shocks presumably merge into the larger-scale
plume flow in a smoother way.

However, the location of the plume base appears to be a function of
the temperature of the host cluster, providing for the first time some
direct evidence that the structure of WATs is determined by their
cluster environments. The model in which the jet-plume transition of
WATs is related to the ISM/ICM interface (which does have the
attractive feature that it `predicts' a scale length for the
transition of tens of kpc, as observed) can account for this only if
the ISM/ICM interface radius is smaller for hotter clusters. We are
not aware of observations that test this prediction at this time, but
in a future paper we will return to this question, \new{and to the
general problem of the physical conditions around the jets and jet
terminations}, using our detailed {\it Chandra} and {\it XMM-Newton}
observations of WAT cluster environments. \newb{Radio observations of
a much larger sample of WATs with similar resolution and sensitivity
would also be valuable to test the generality of our conclusions.}

\section*{Acknowledgements}

The National Radio Astronomy Observatory is a facility of the National
Science Foundation operated under cooperative agreement by Associated
Universities, Inc. The Australia Telescope is funded by the
Commonwealth of Australia for operation as a National Facility managed
by CSIRO. We are grateful to Jean Eilek and Debora Katz-Stone for
allowing us to use their archival VLA data, to the staff of the ATNF,
particularly Tasso Tzioumis, for their help in carrying out the ATCA
observations and in organizing our observing visits, and to Nazirah
Jetha for discussion of WAT properties. In collating X-ray data for
the clusters we made use of the BAX (Base de donn\'ees Amas de
galaxies X) system at http://webast.ast.obs-mip.fr/bax/ . \new{We
thank an anonymous referee for comments that helped us to improve the
paper.}

MJH thanks the Royal Society for a Research Fellowship. MJH and IS
would like to thank the Universities of Birmingham and Bristol,
respectively, for hospitality during the preparation of this paper.
\new{IS is grateful to the Mullard Space Science Laboratory for funding
during the early stages of this work.}

\appendix
\section{The jets of 3C\,75}

\new{The well-known twin source 3C\,75 (Owen \etal\ 1985), at a
  redshift of 0.0240, is often classed as a wide-angle tail on the
  basis of its large-scale structure (e.g. O'Donoghue \etal\ 1993). It
  is certainly the case that the object lies at the centre of a
  cluster, Abell 400, and that its large-scale plumes bend in a manner
  similar to those of some of the WATs in our sample, \newb{although the
  radio luminosities of the two sources are a factor $\sim 5$ lower
  than the typical luminosity of our WATs, and much more comparable to
  those of normal twin-jet FRIs}. However, we have excluded it from
  consideration here on the basis that its inner jets have a large
  opening angle and that it does not show an abrupt jet-plume
  transition with a bright plume base; \newb{we would instead describe it as
  a pair of normal twin-jet FRI sources in a rich environment}. To
  justify this, we present here a previously unpublished
  high-resolution VLA image of the inner parts of the source. The
  source was observed by one of us (MJH) at two frequencies, 8.415 and
  8.465 GHz, with a 50-MHz bandwidth for 2 hours in B-configuration on
  1995 Nov 28 and 1 hour in C-configuration on 1994 Nov 10. The data
  were reduced as described in Section \ref{radio}. Fig. \ref{3c75}
  shows the resulting image. It is clear that the jets of 3C\,75 are
  well resolved, with an opening angle much larger than those of the
  WATs in our sample (compare 3C\,465, at a similar redshift) \newb{and have
  no strong side-to-side asymmetry, as is typical for normal FRI jets
  at more than a few kpc from the core}. The N jet also displays the
  transition between parallel and perpendicular magnetic field
  characteristic of normal FRI jets, which takes place at a distance
  from the core about 8 arcsec (4 kpc). The fainter S jets have a
  perpendicular magnetic field from about 16 arcsec (8 kpc). There is
  no bright flare point or abrupt change in opening angle in these
  images, or in the lower-resolution but more sensitive images of Owen
  \etal\ (1985) or Eilek \& Owen (2002); instead, the jets merge
  seamlessly into the plumes. We thus feel justified in excluding this
  object from our sample of WATs, although it is present in the parent
  sample of bent sources (Sakelliou \& Merrifield 2000).}

\bsp

\renewcommand{\thefigure}{\arabic{figure}}
\begin{figure*}
\epsfxsize 8cm
\epsfbox{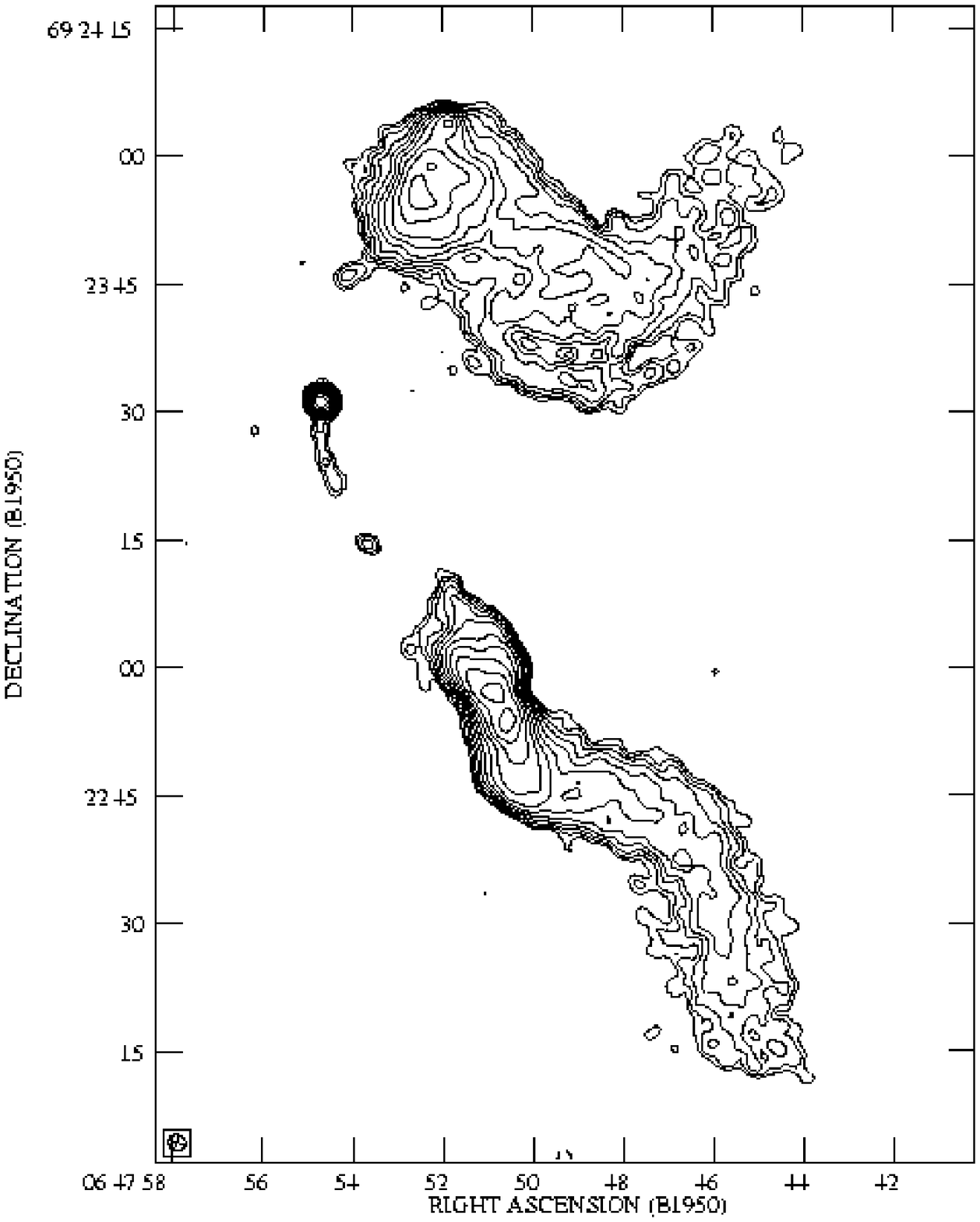}
\epsfxsize 8cm
\epsfbox{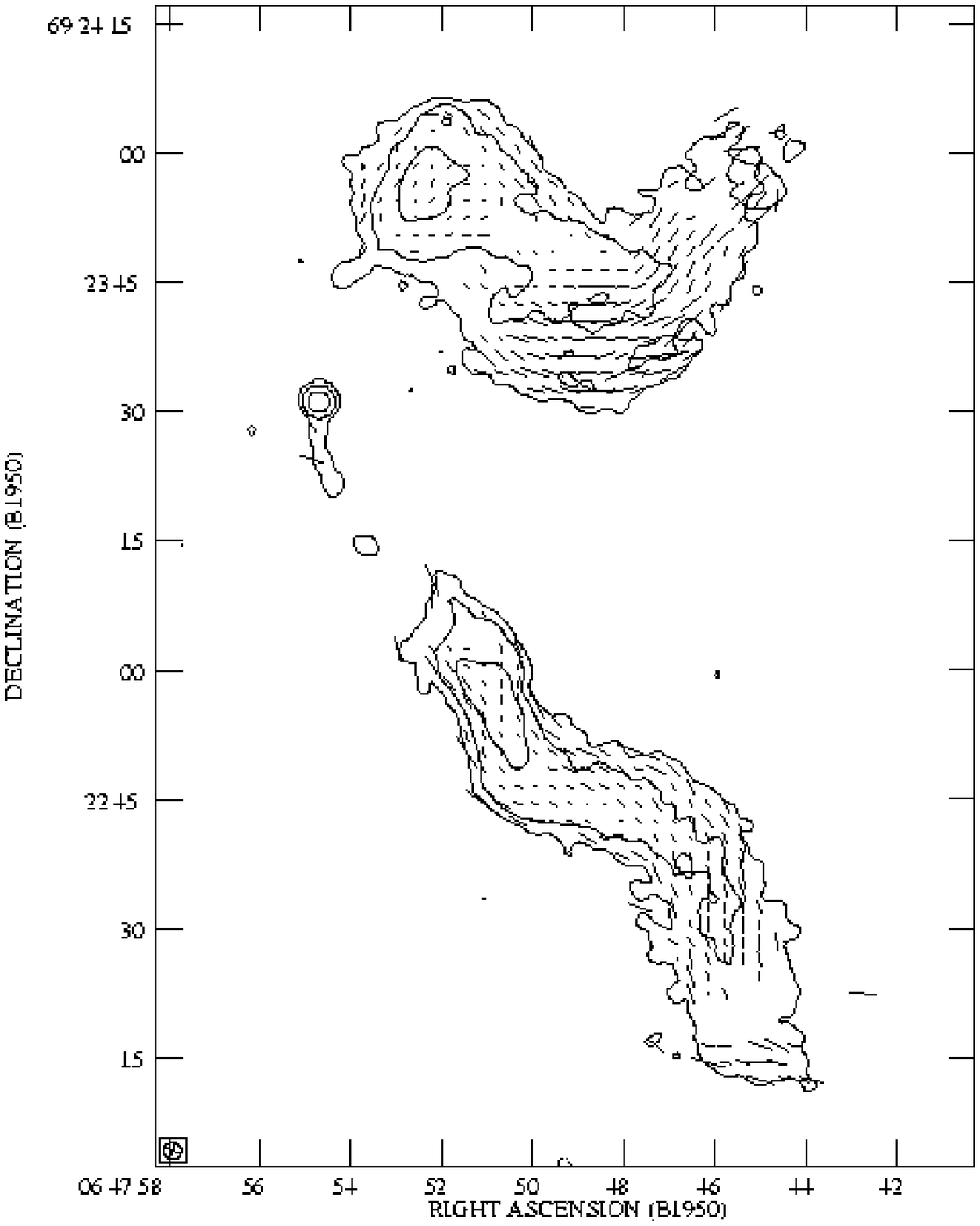}
\caption{2-arcsec resolution map of 0647+693. The left-hand map shows
the total intensity: contours are logarithmic in steps of $\sqrt{2}$,
with the base contour being 100 $\mu$Jy beam$^{-1}$. Negative contours
are dashed. The right-hand
map shows polarization: logarithmic total intensity contours with the
same base level, but increasing by a factor 4, are plotted on top of a
vector map. Vector magnitudes show the fractional polarization while
the direction is that of the $B$-field, as discussed in the text.}
\label{0647l}
\end{figure*}

\begin{figure*}
\epsfxsize 10cm
\epsfbox{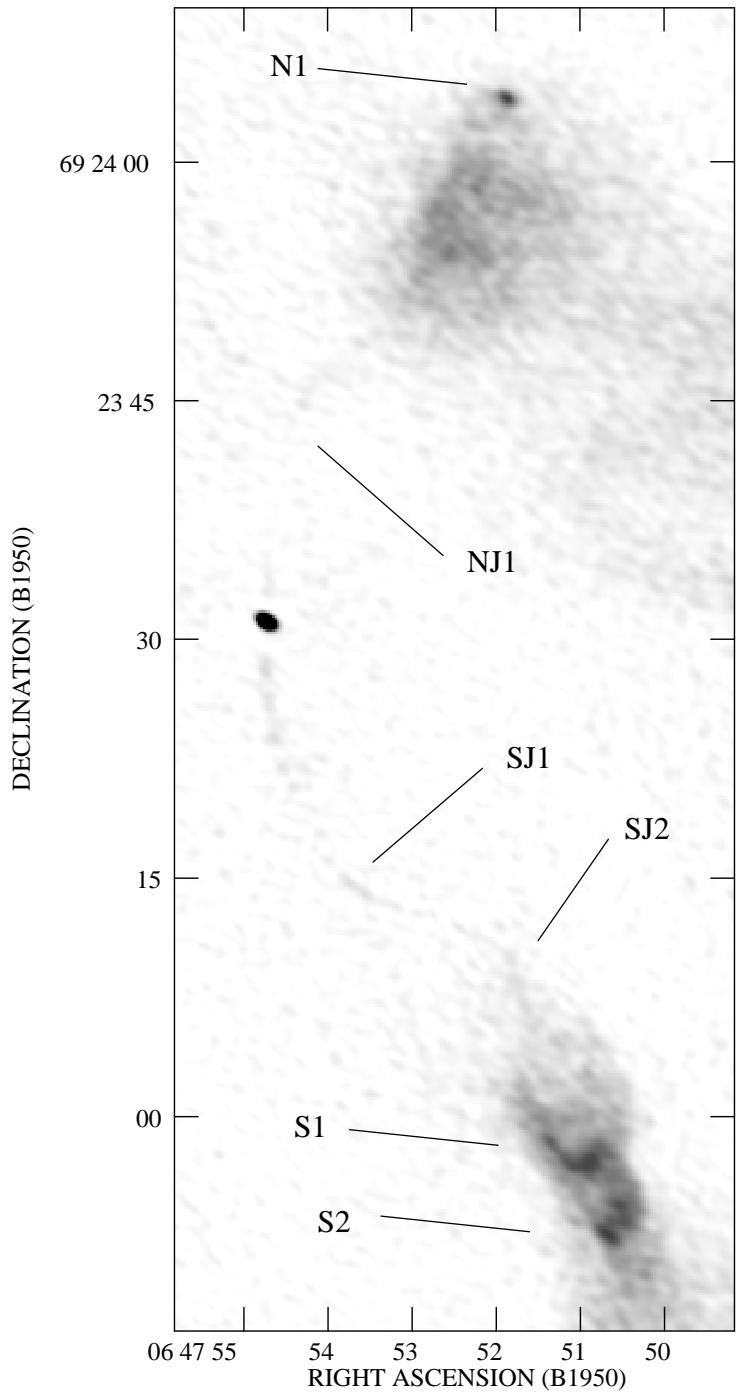}
\caption{$0.9 \times 0.6$-arcsec resolution map of 0647+693. Black is
1 mJy beam$^{-1}$.}
\label{0647h}
\end{figure*}

\begin{figure*}
\epsfxsize 8cm
\epsfbox{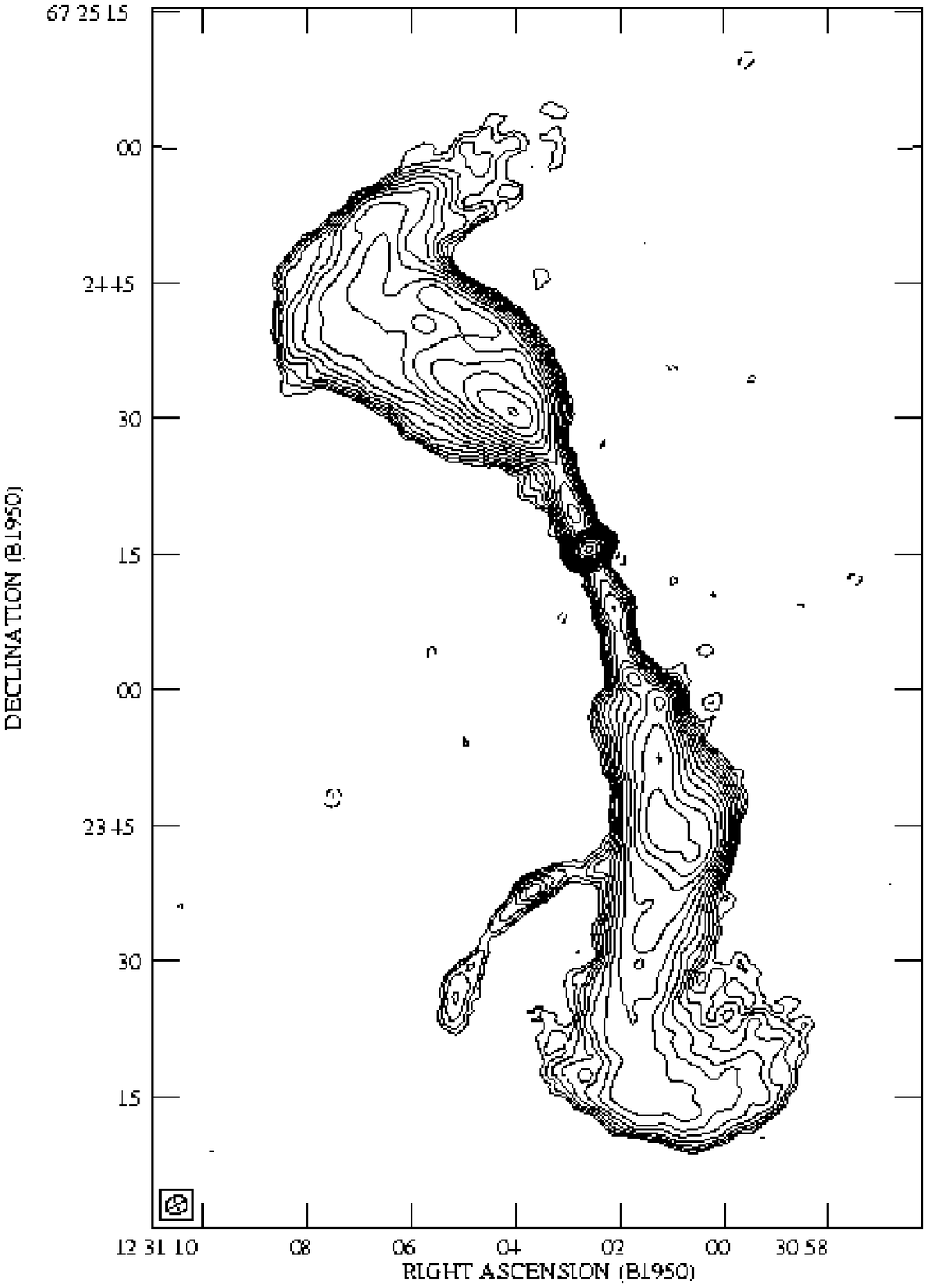}
\epsfxsize 8cm
\epsfbox{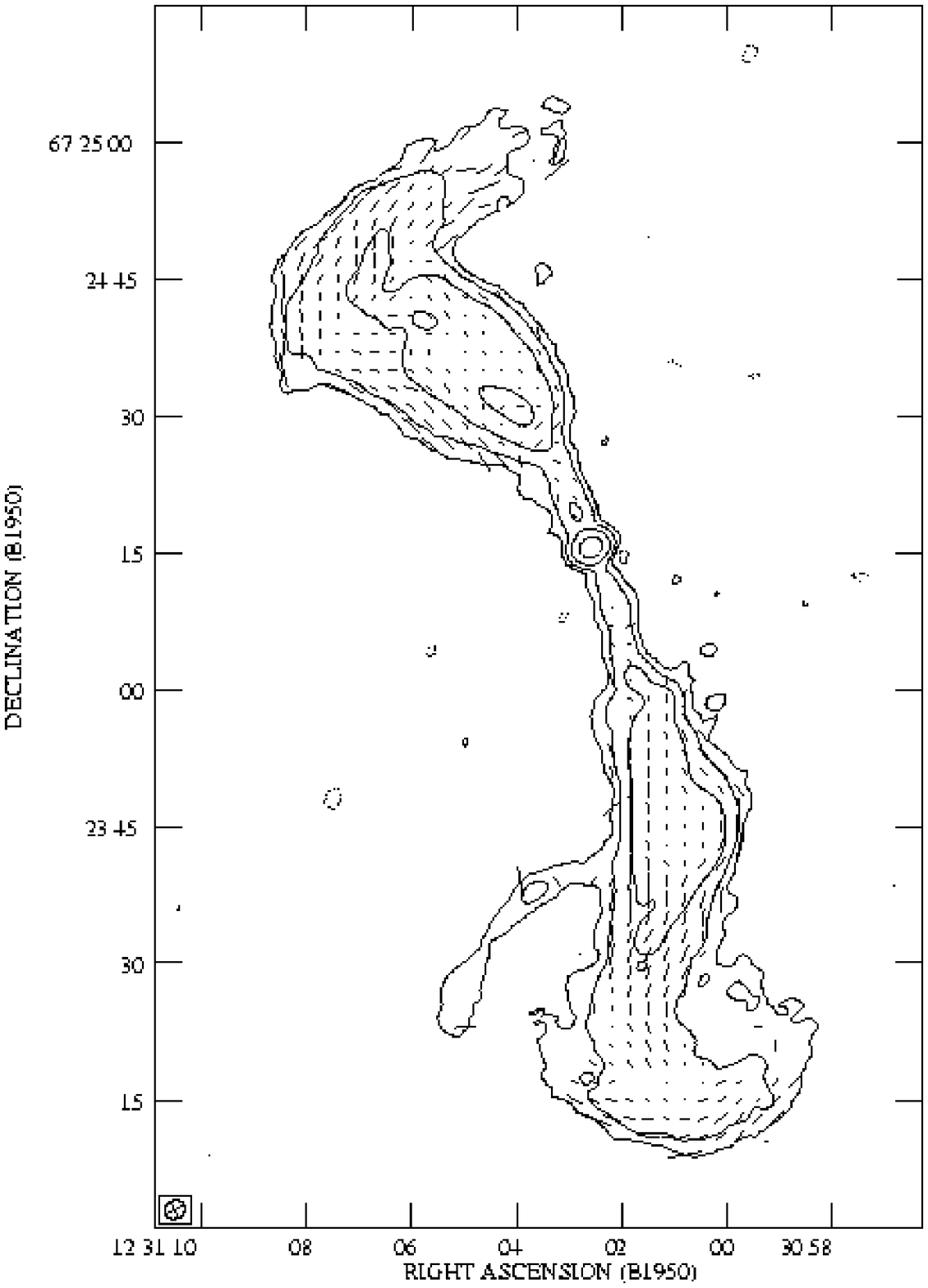}
\caption{2-arcsec resolution map of 1231+674. As Fig.\ \ref{0647l},
but lowest contour level is 40 $\mu$Jy beam$^{-1}$.}
\label{1231l}
\end{figure*}

\begin{figure*}
\epsfxsize 8.0cm
\epsfbox{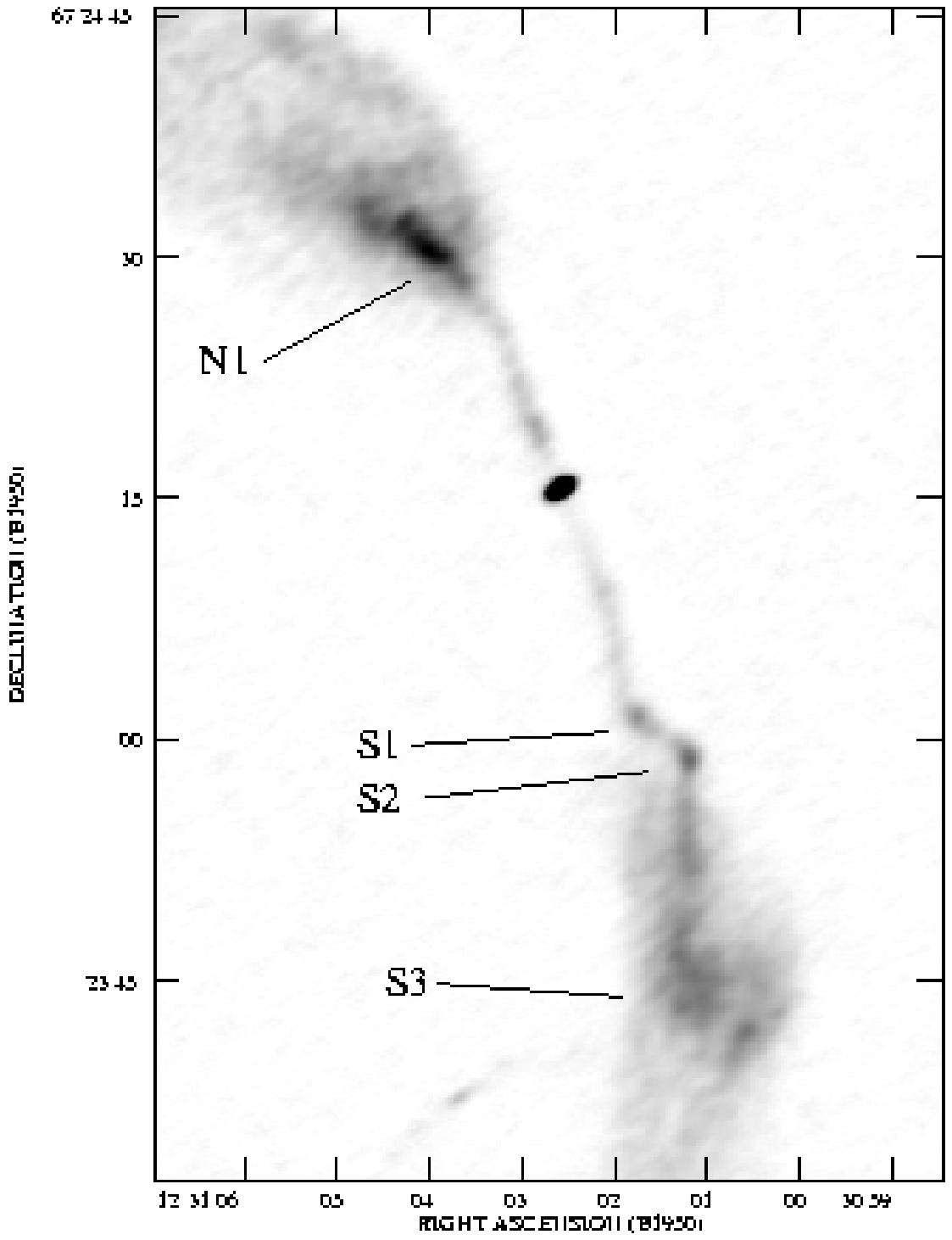}
\epsfxsize 8.0cm
\epsfbox{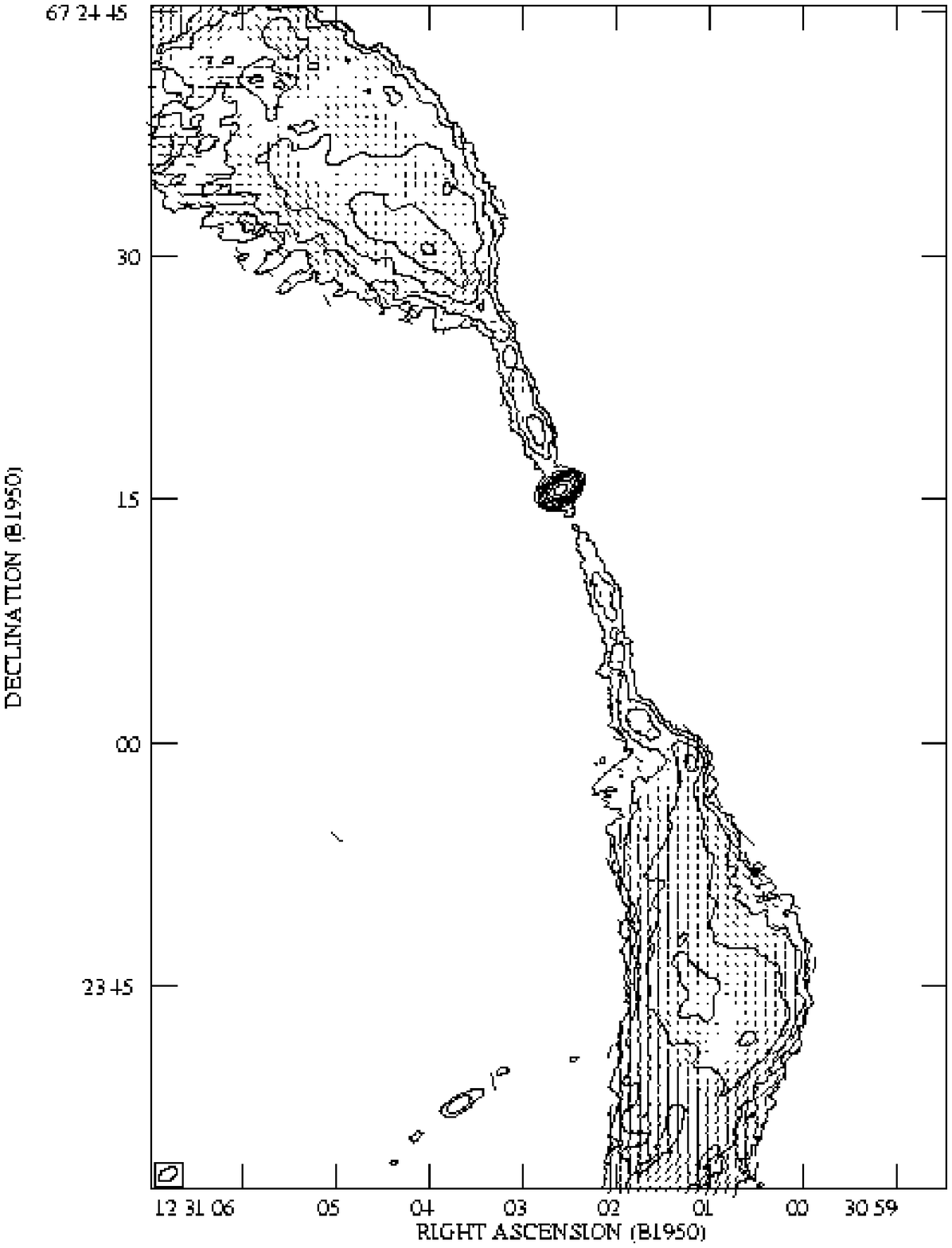}
\caption{$1.24 \times 0.73$-arcsec resolution map of 1231+674. Left: black
is 1 mJy beam$^{-1}$. Right: As Fig.\ \ref{0647l} (right),
but lowest contour level is 60 $\mu$Jy beam$^{-1}$.}
\label{1231h}
\end{figure*}

\begin{figure*}
\epsfxsize 11cm
\epsfbox{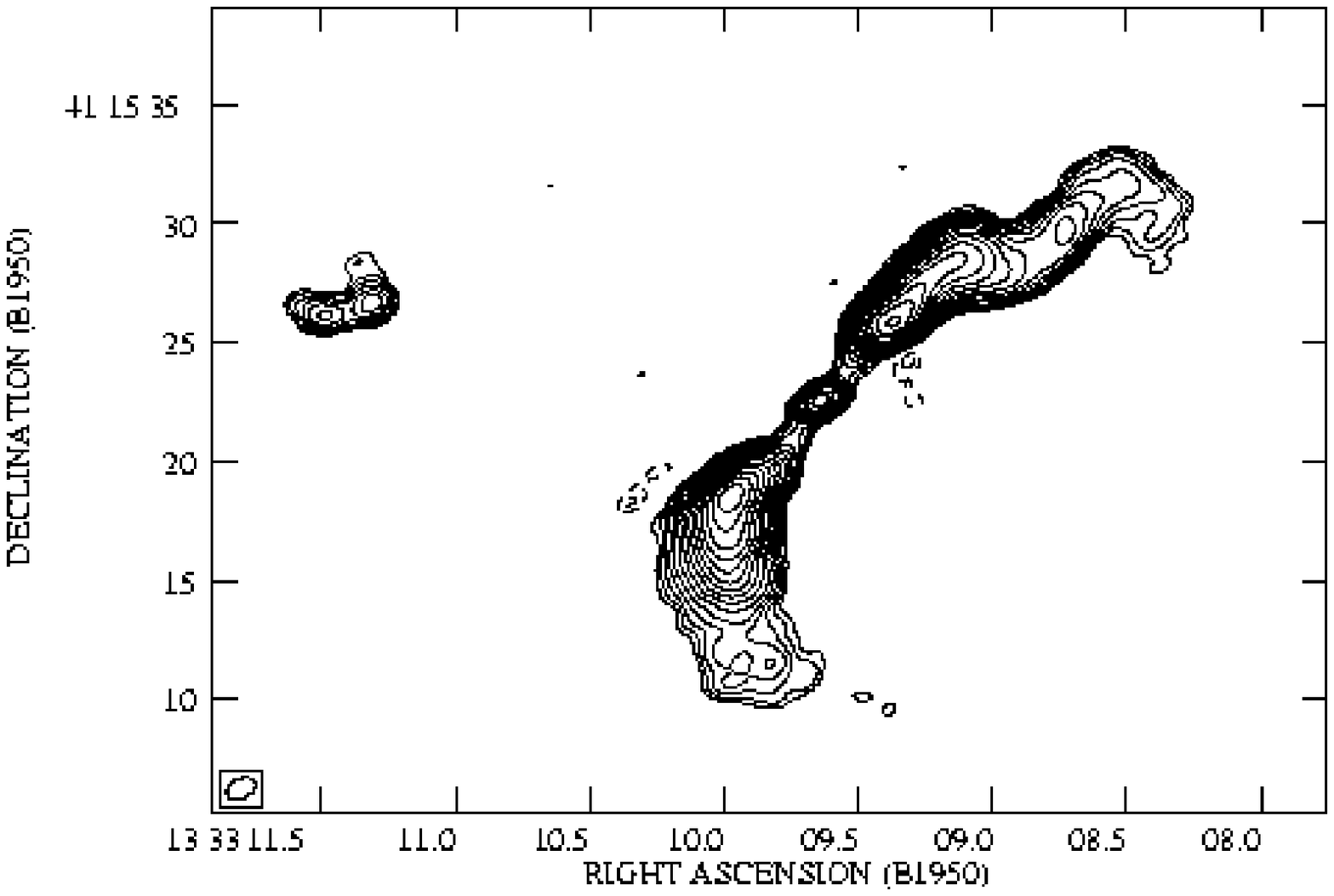}
\epsfxsize 11cm
\epsfbox{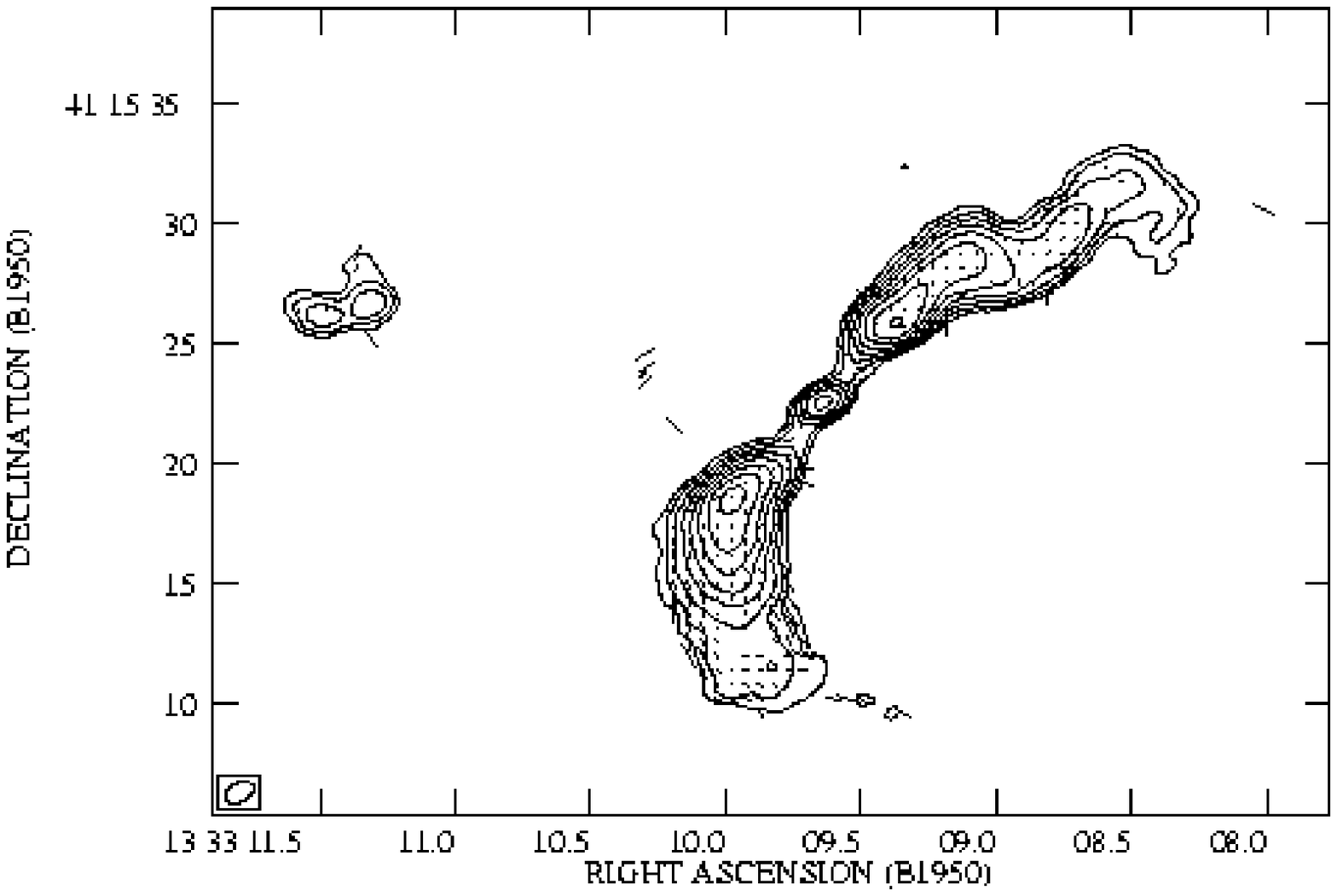}
\caption{$1.3 \times 0.8$-arcsec resolution map of 1333+412. As Fig.\
\ref{0647l}, but lowest contour level is 80 $\mu$Jy beam$^{-1}$.}
\label{1333l}
\end{figure*}

\begin{figure*}
\epsfxsize 11cm
\epsfbox{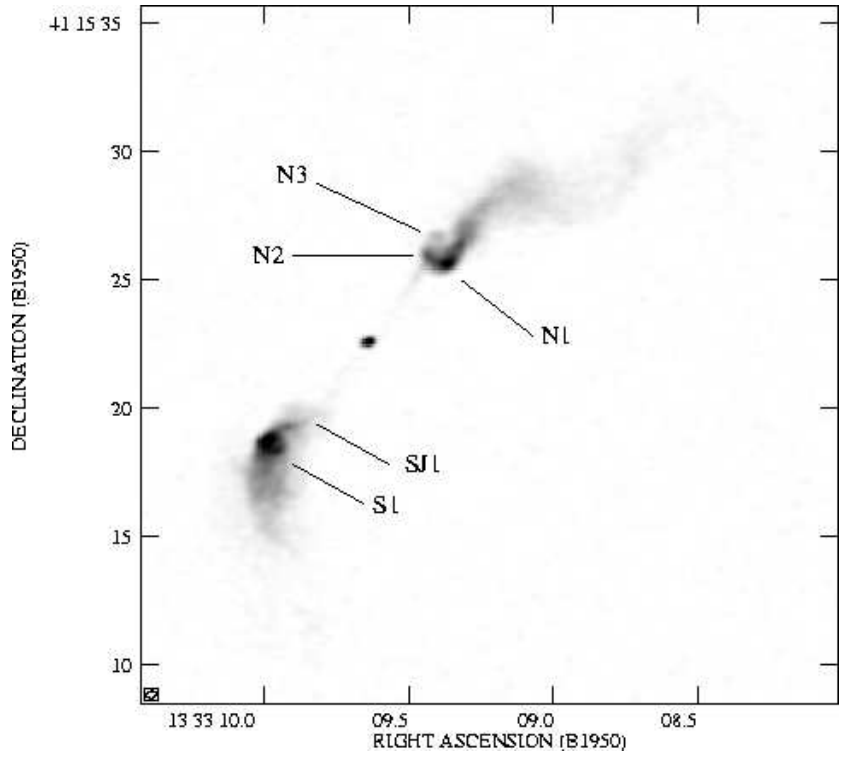}
\caption{$0.38 \times 0.27$-arcsec resolution map of 1333+412. Black is
2 mJy beam$^{-1}$.}
\label{1333h}
\end{figure*}
\begin{figure*}
\epsfxsize 8cm
\epsfbox{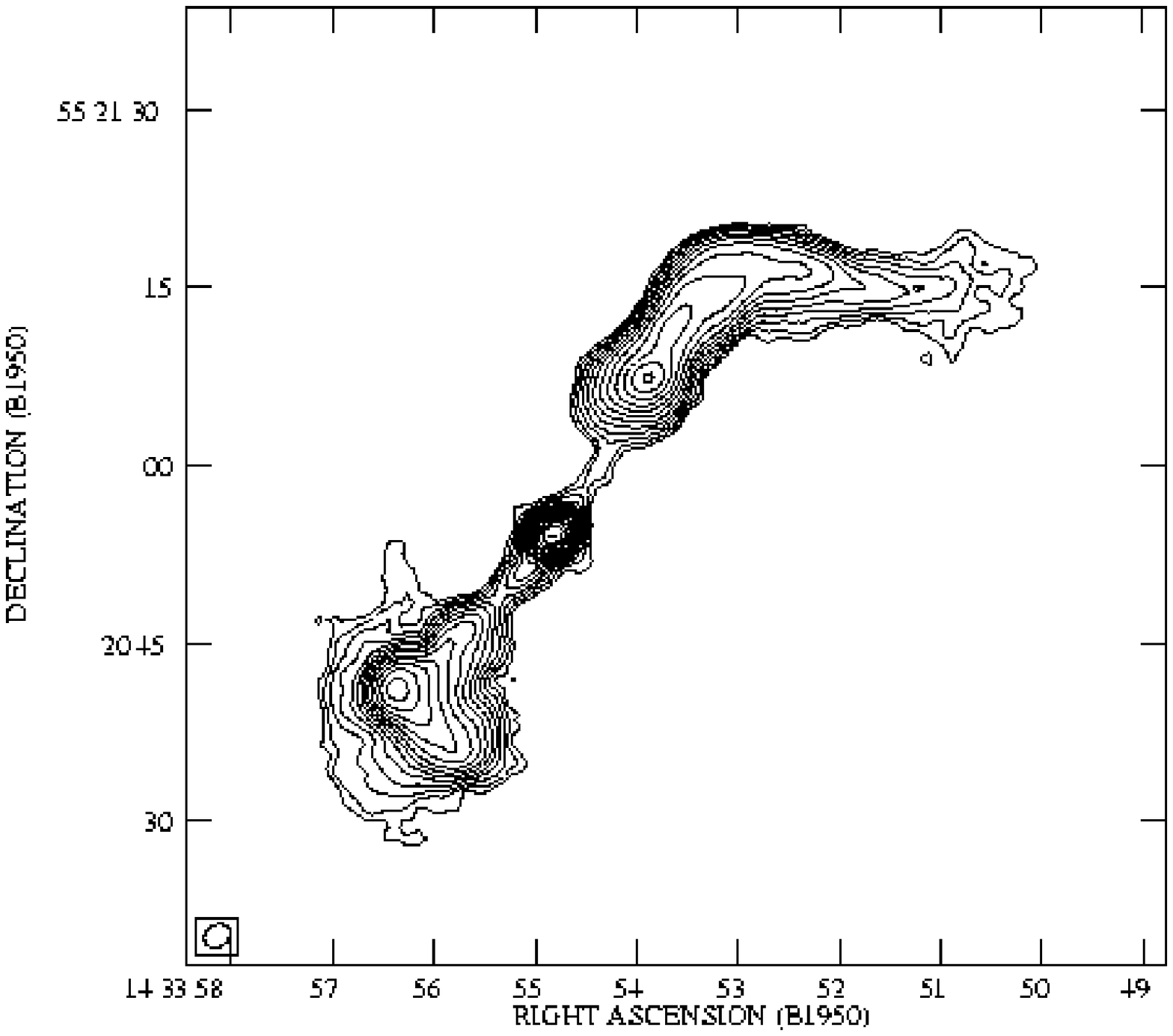}
\epsfxsize 8cm
\epsfbox{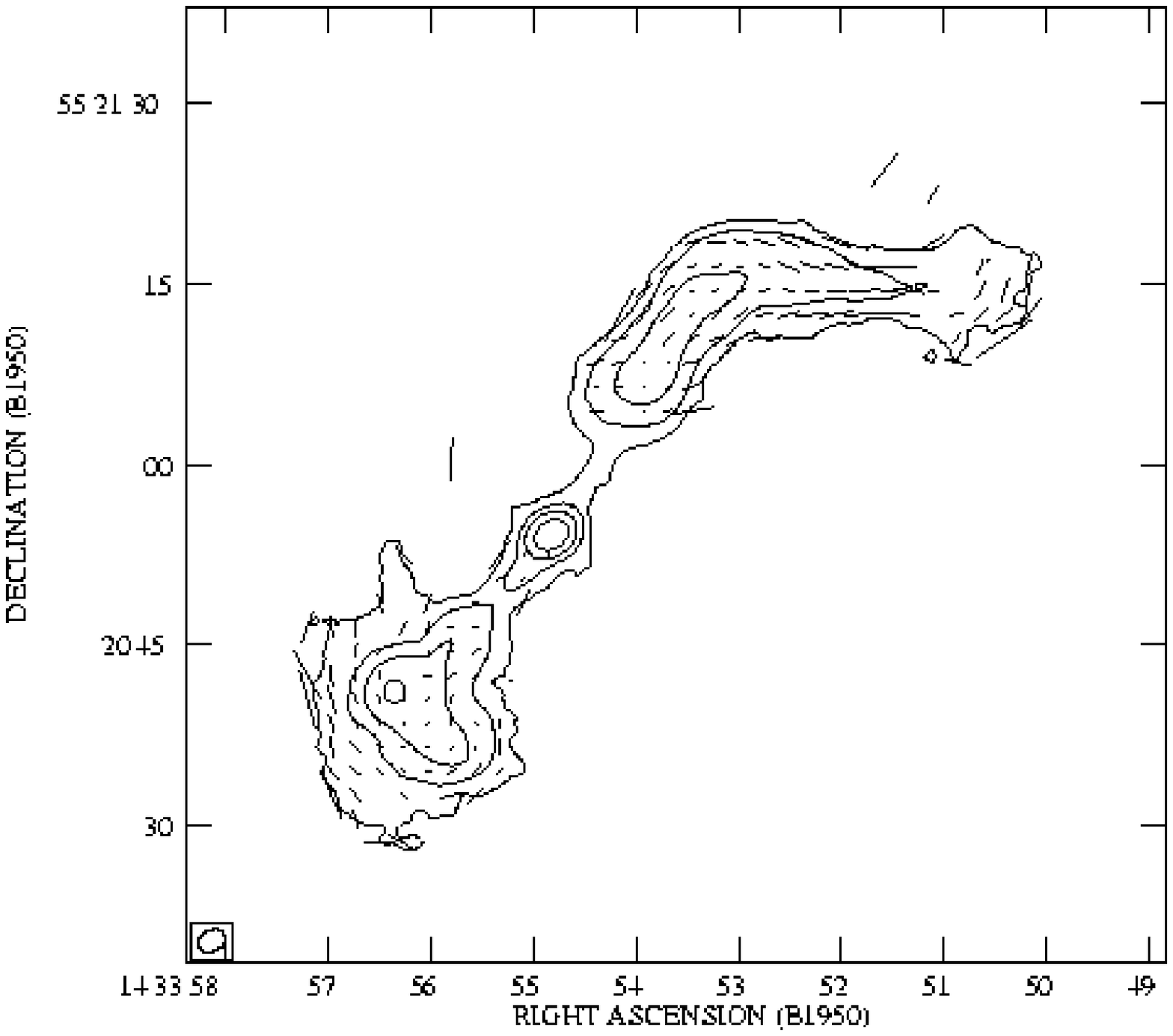}
\caption{$2.2 \times 1.8$-arcsec resolution map of 1433+553. As Fig.\
\ref{0647l}, but lowest contour level is 80 $\mu$Jy beam$^{-1}$.}
\label{1433l}
\end{figure*}

\begin{figure*}
\epsfxsize 11cm
\epsfbox{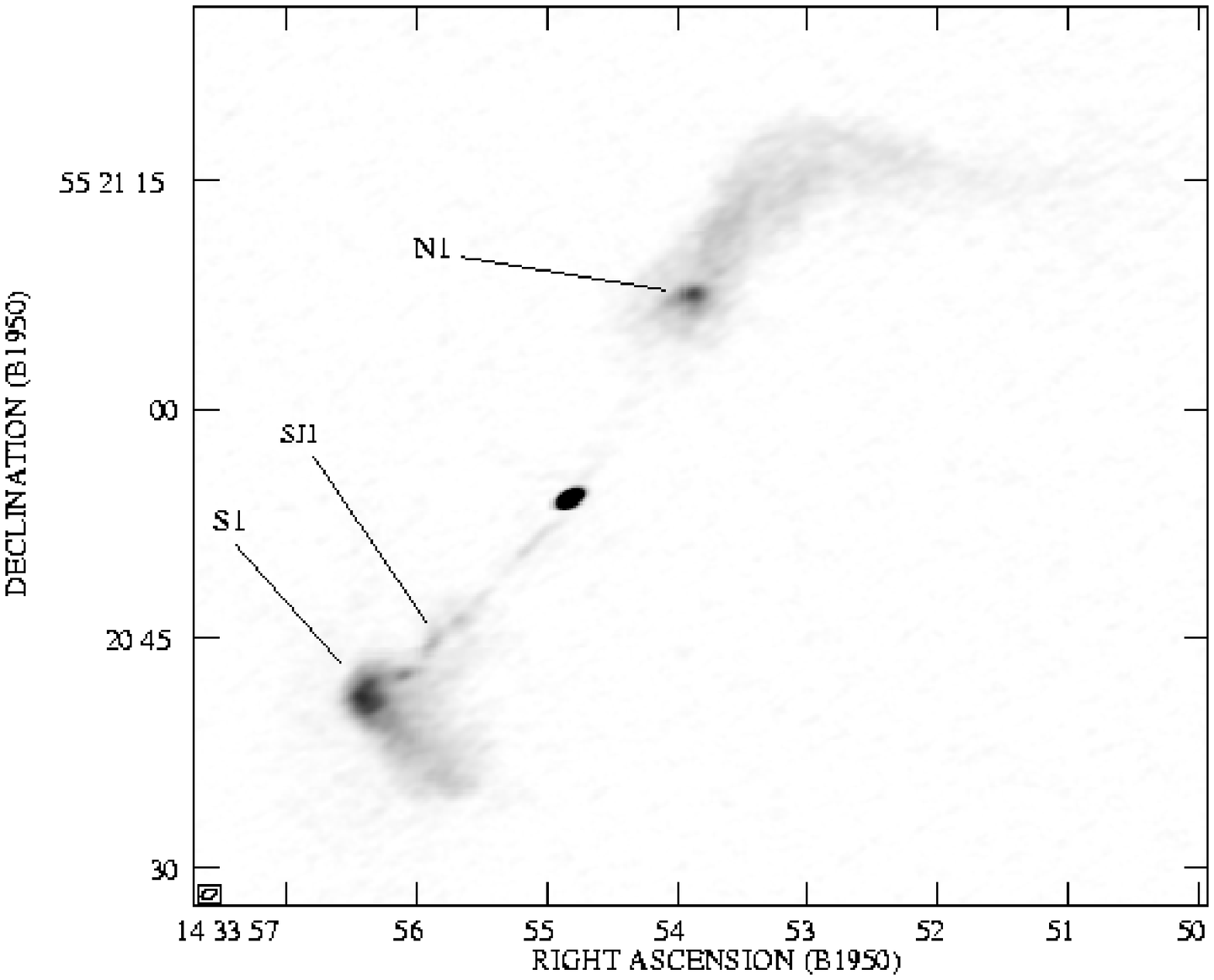}
\epsfxsize 11cm
\epsfbox{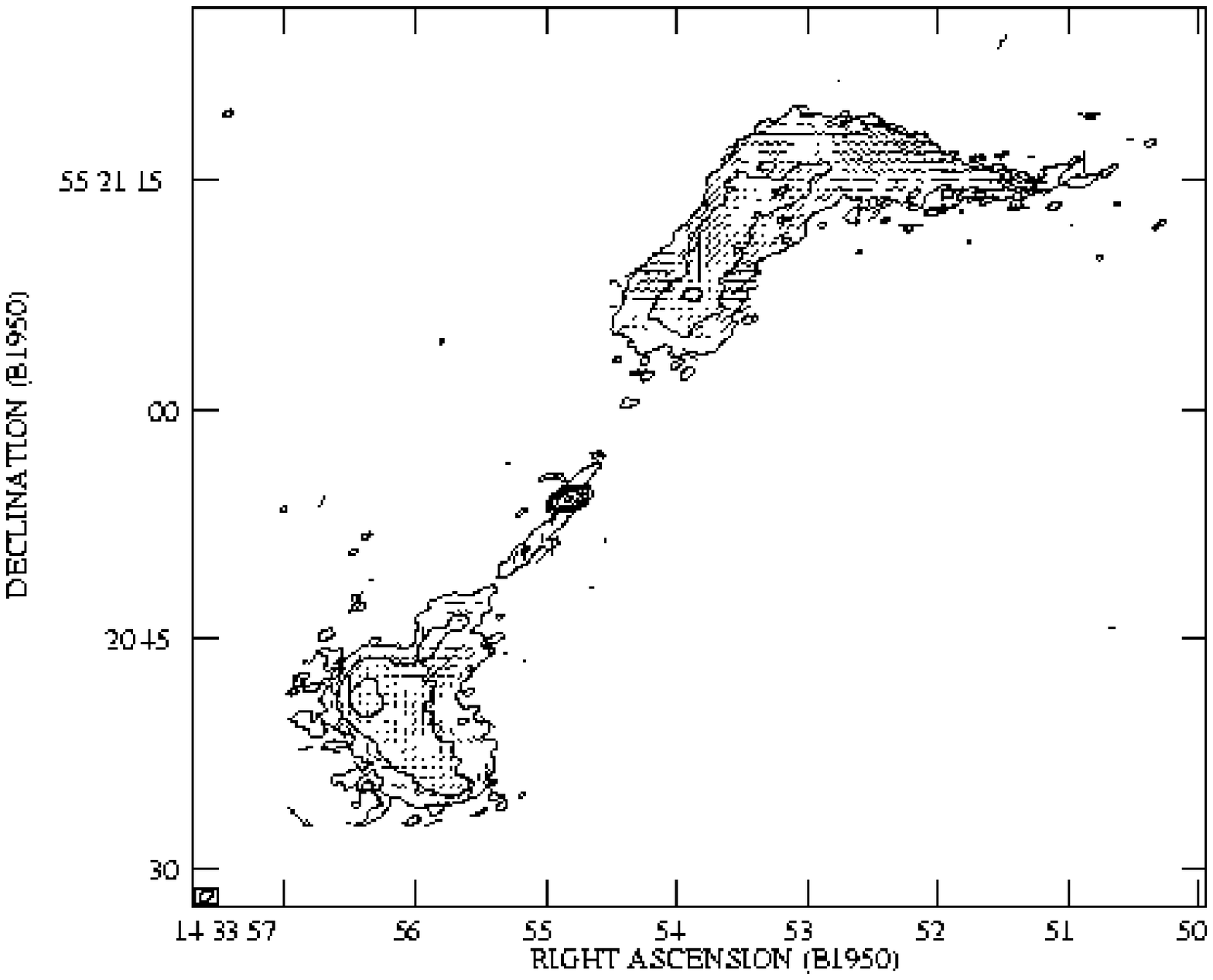}
\caption{$1.0 \times 0.55$-arcsec resolution maps of 1433+553. Top:
  black is 1.5 mJy beam$^{-1}$. Bottom: as Fig.\ \ref{0647l}, but
  lowest contour is 100 $\mu$Jy beam$^{-1}$.}
\label{1433h}
\end{figure*}

\begin{figure*}
\epsfxsize 8cm
\epsfbox{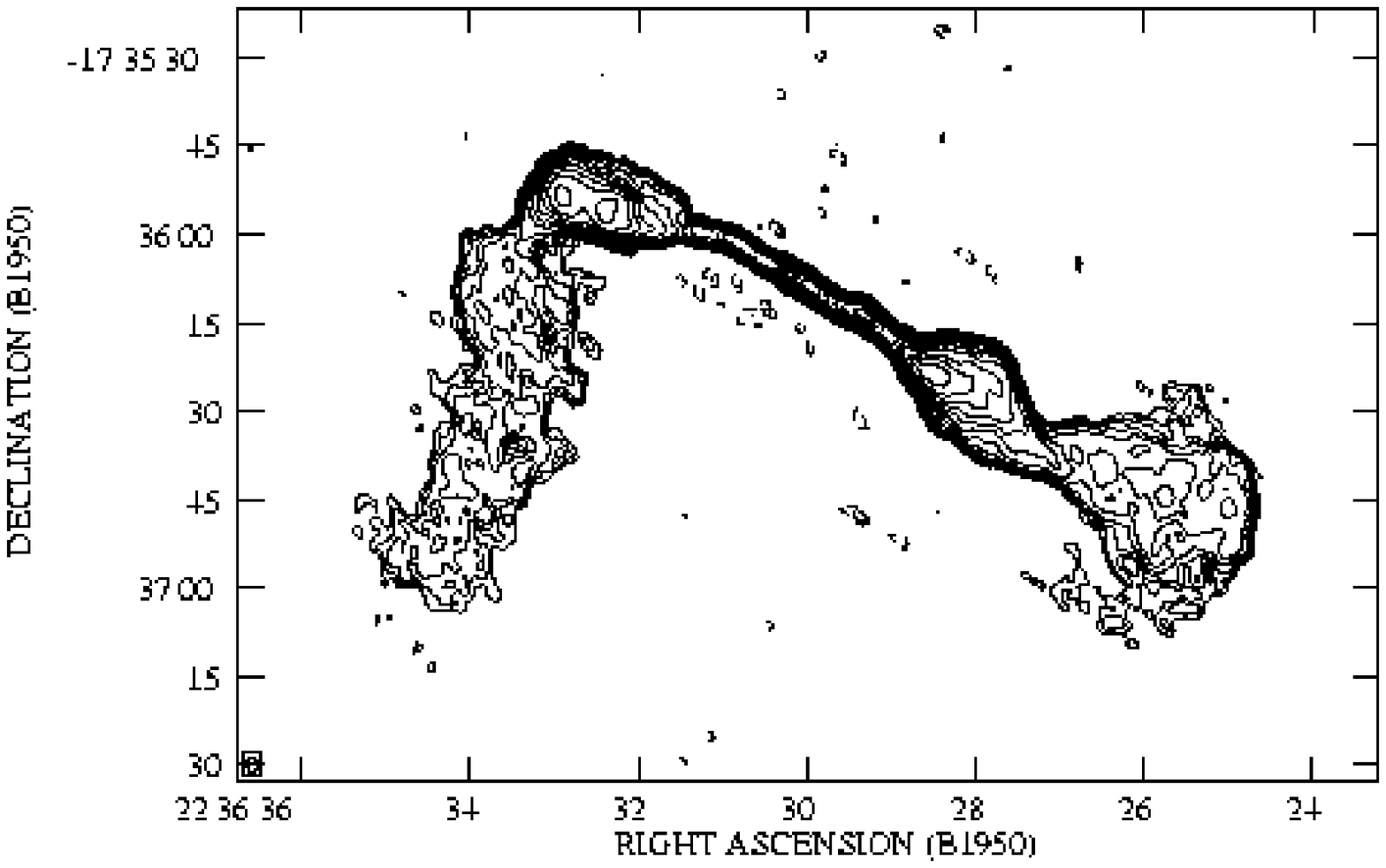}
\epsfxsize 8cm
\epsfbox{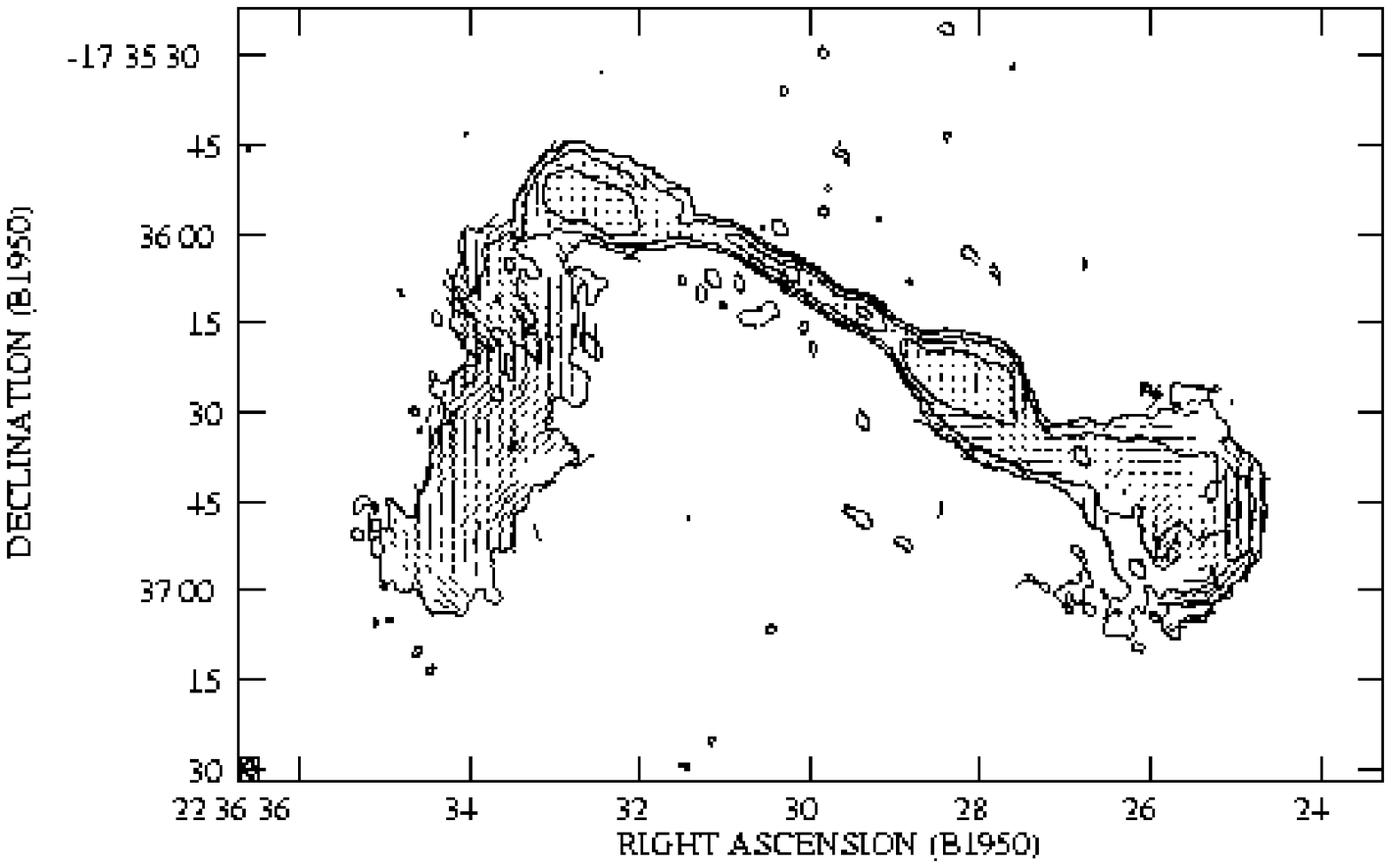}
\caption{$2.6 \times 1.8$-arcsec resolution map of 2236$-$176. As Fig.\
\ref{0647l}, but lowest contour level is 80 $\mu$Jy beam$^{-1}$.}
\label{2236l}
\end{figure*}

\begin{figure*}
\epsfxsize 13cm
\epsfbox{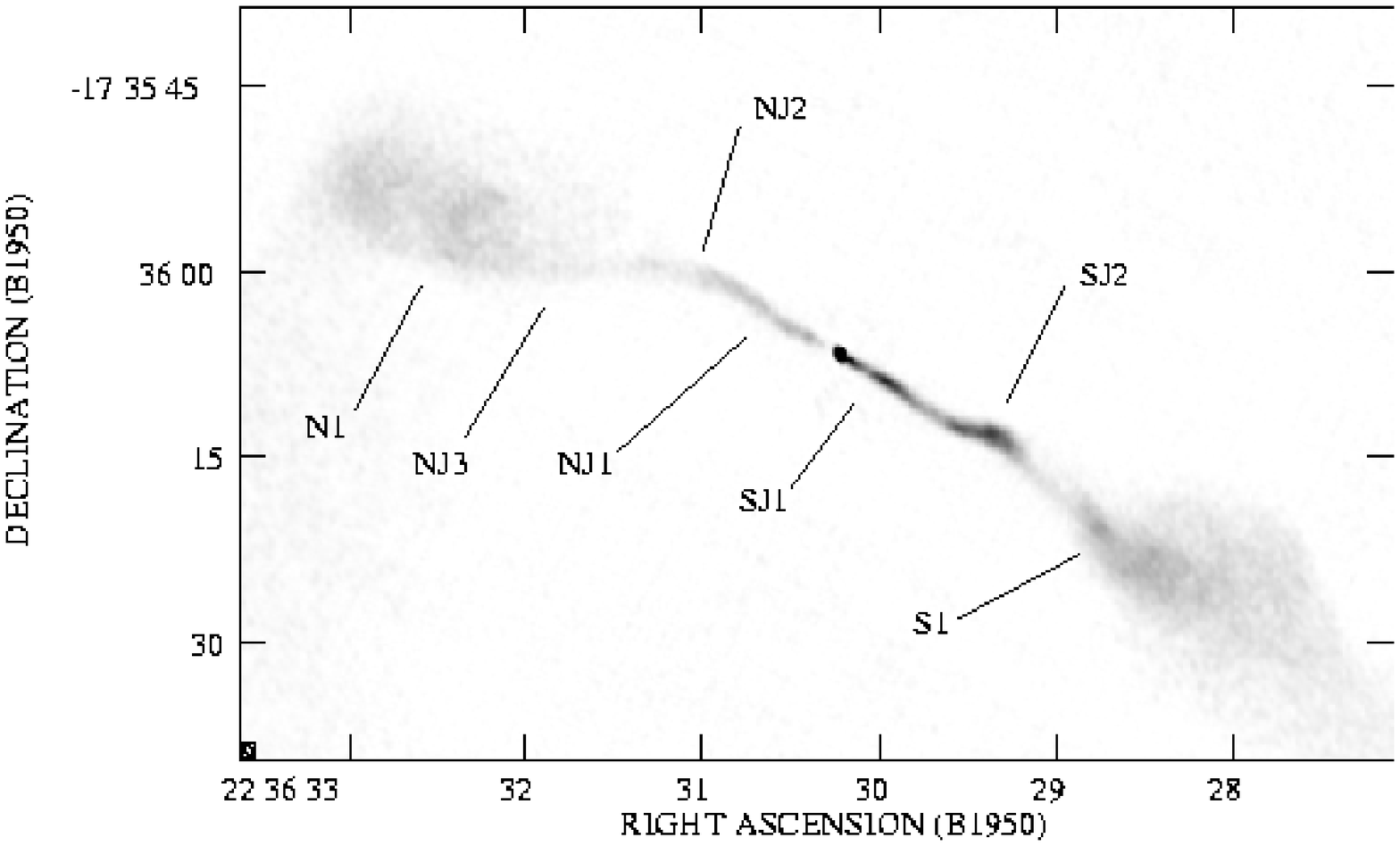}
\epsfxsize 13cm
\epsfbox{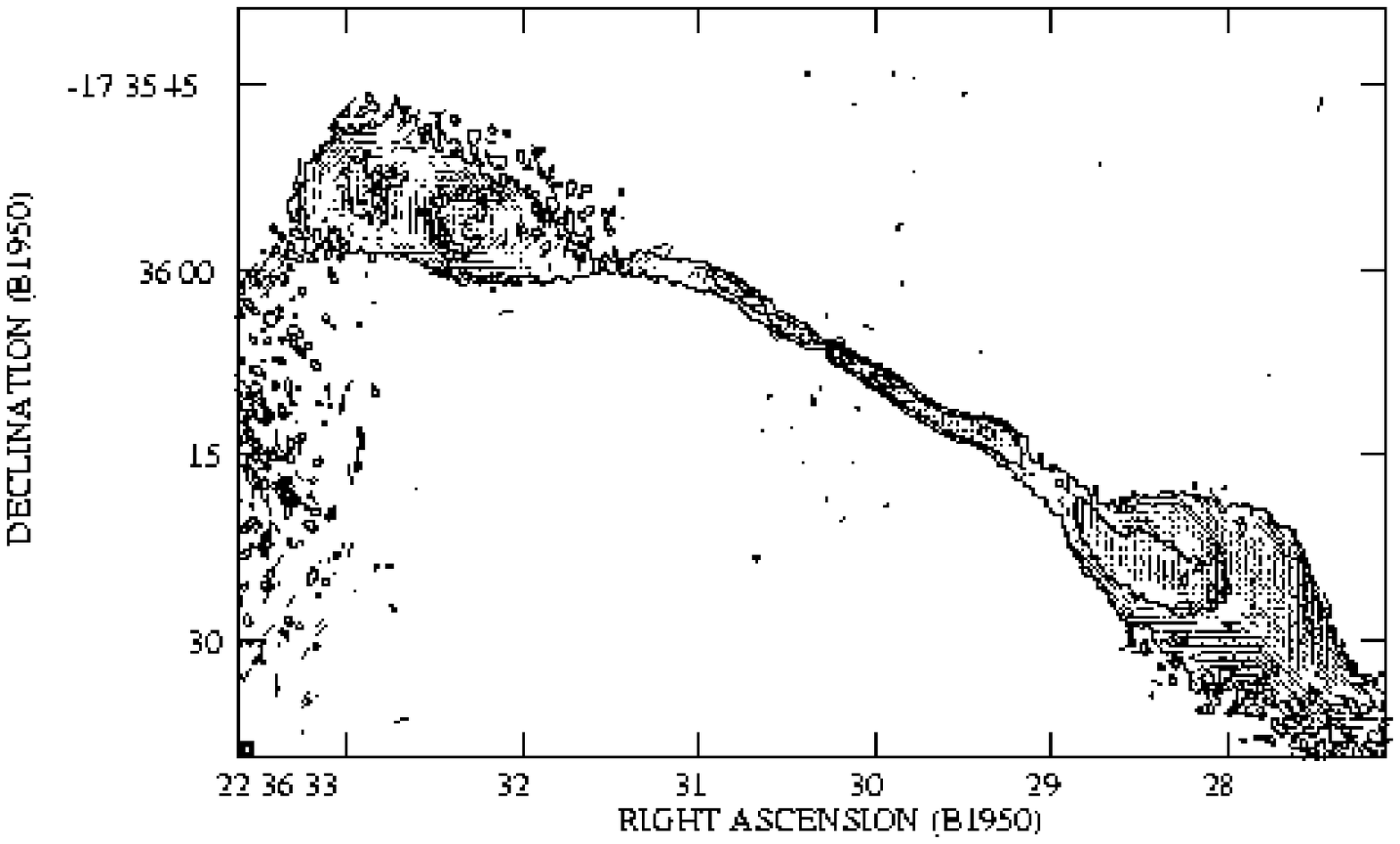}
\caption{$0.78 \times 0.52$-arcsec resolution map of 2236$-$176. Top: black
is 1.5 mJy beam$^{-1}$. Bottom: As Fig.\ \ref{0647l} (right),
but lowest contour level is 60 $\mu$Jy beam$^{-1}$.}
\label{2236h}
\end{figure*}

\begin{figure*}
\epsfxsize 12cm
\epsfbox{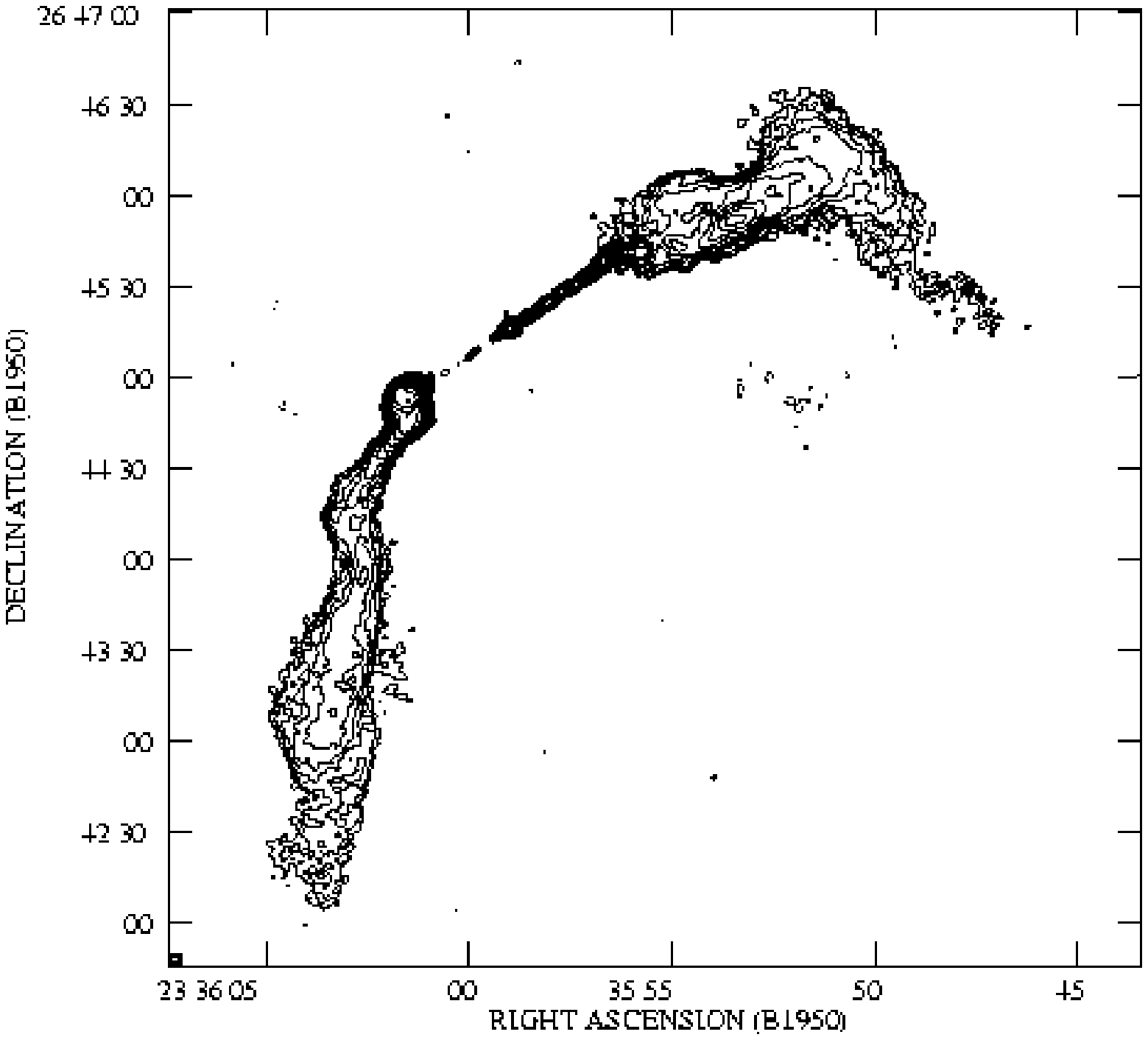}
\epsfxsize 12cm
\epsfbox{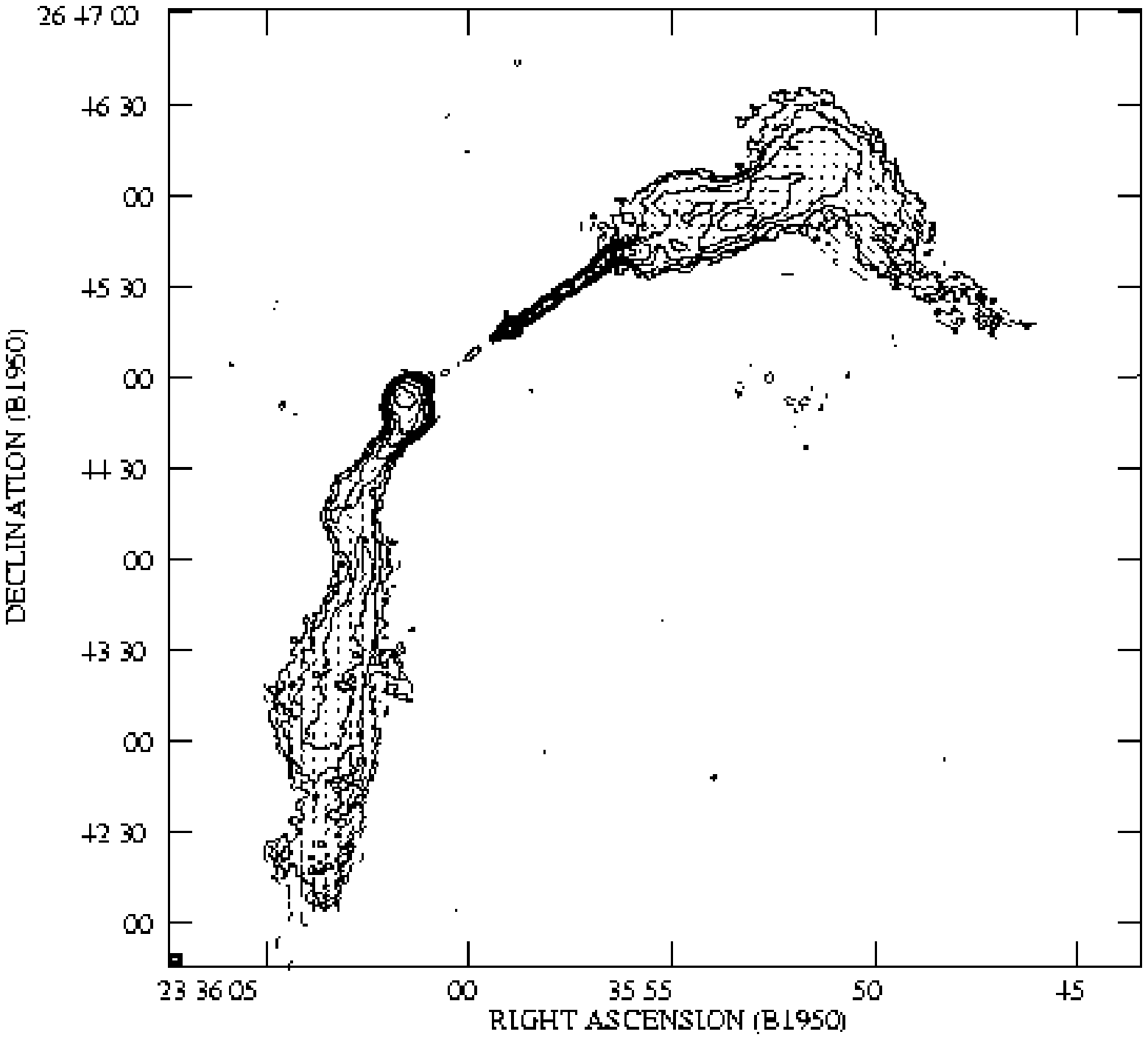}
\caption{$2.6 \times 2.0$-arcsec resolution map of 3C\,465. As Fig.\
\ref{0647l}, but lowest contour level is 130 $\mu$Jy beam$^{-1}$.}
\label{465l}
\end{figure*}

\begin{figure*}
\epsfxsize 12cm
\epsfbox{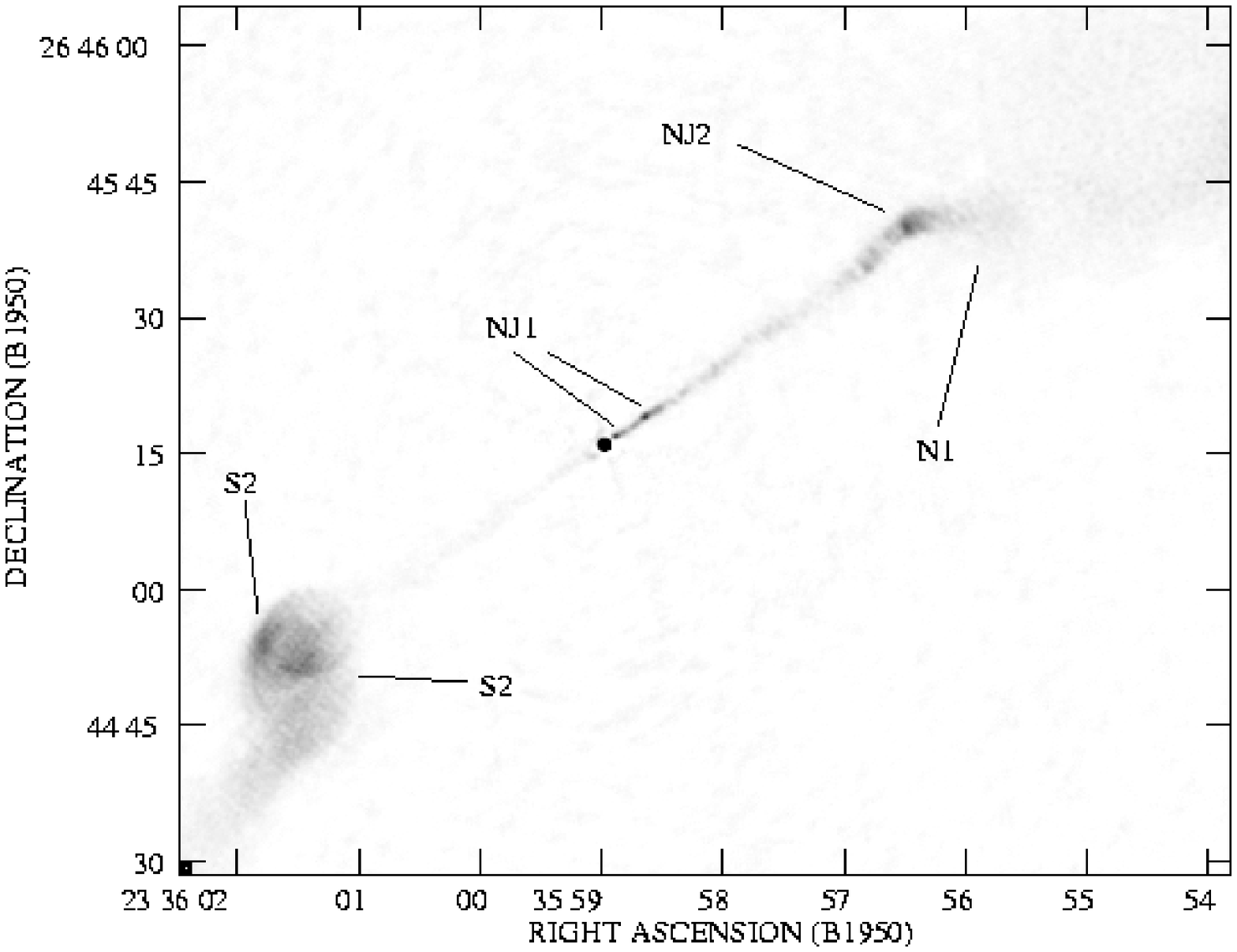}
\hbox{
\epsfysize 7cm
\epsfbox{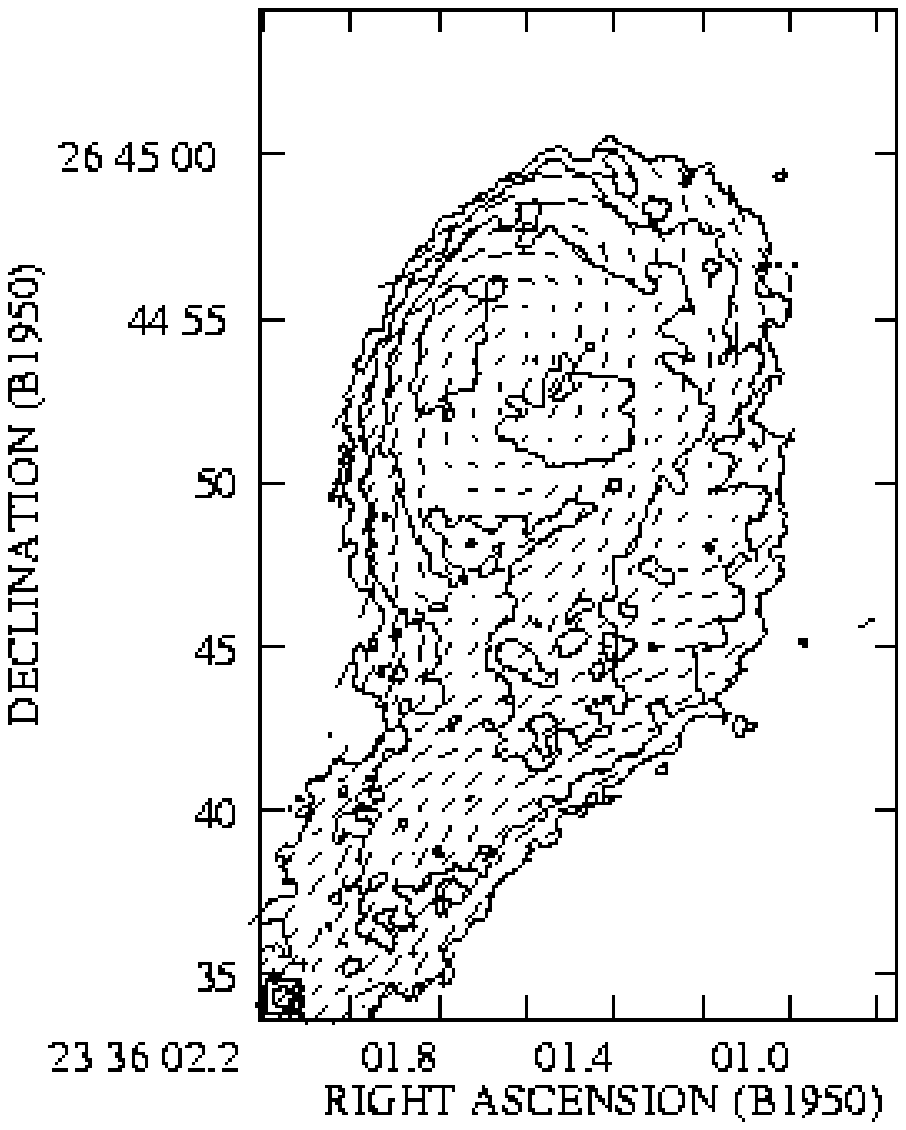}
\epsfysize 7cm
\epsfbox{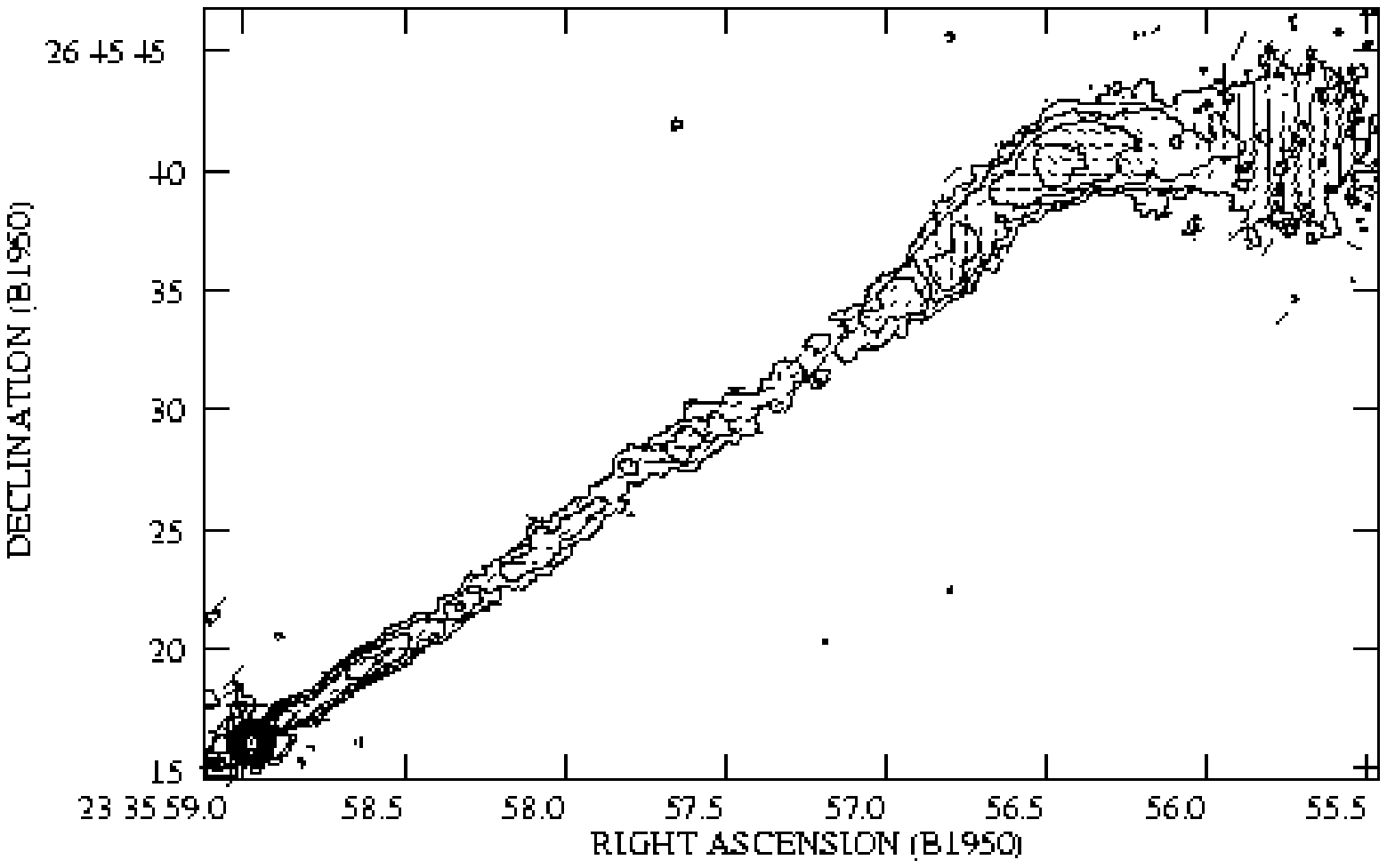}
}
\caption{$0.61 \times 0.60$-arcsec resolution maps of 3C\,465. Top:
black is 2 mJy beam$^{-1}$. Bottom: as Fig.\ \ref{0647l}, but lowest
contour level is 100 $\mu$Jy beam$^{-1}$. The two polarization maps
show the structure in (left) the S plume base and (right) the jet.}
\label{465h}
\end{figure*}

\begin{figure*}
\epsfxsize 13cm
\epsfbox{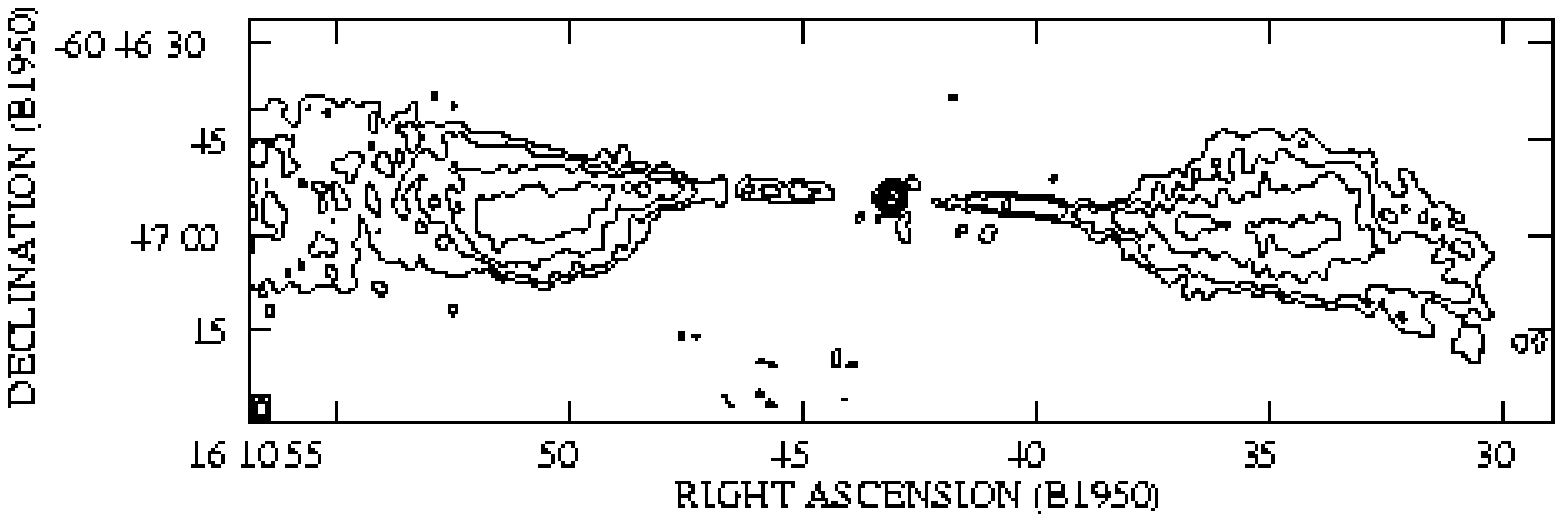}
\epsfxsize 13cm
\epsfbox{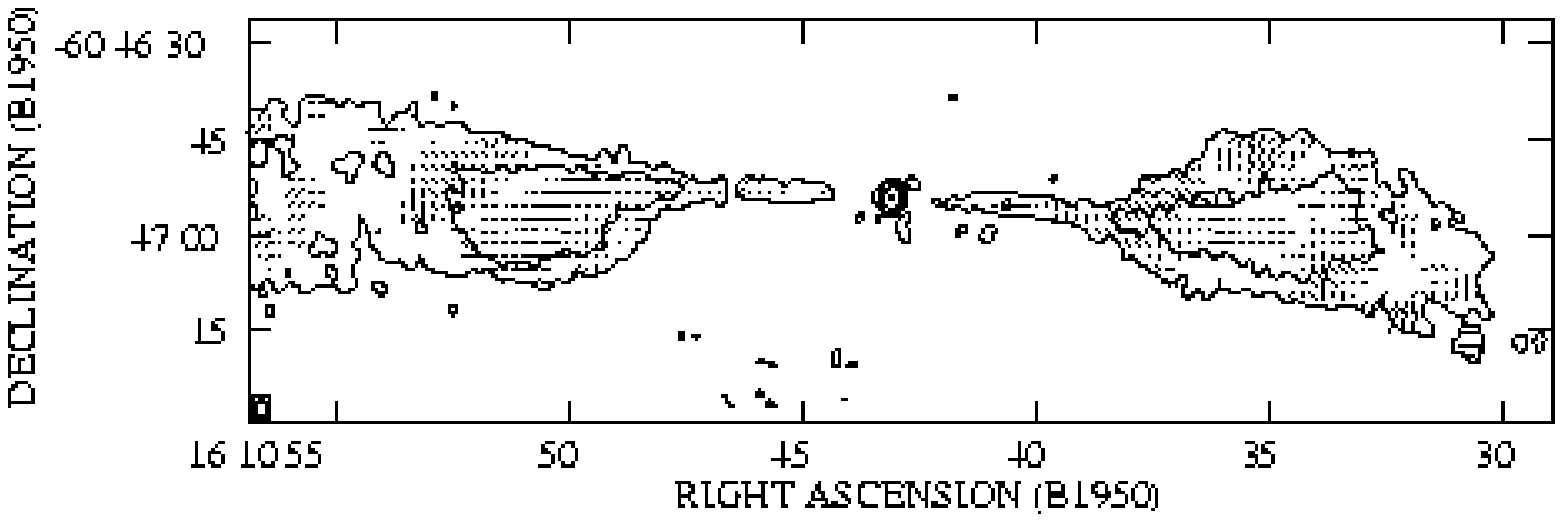}
\epsfxsize 13cm
\epsfbox{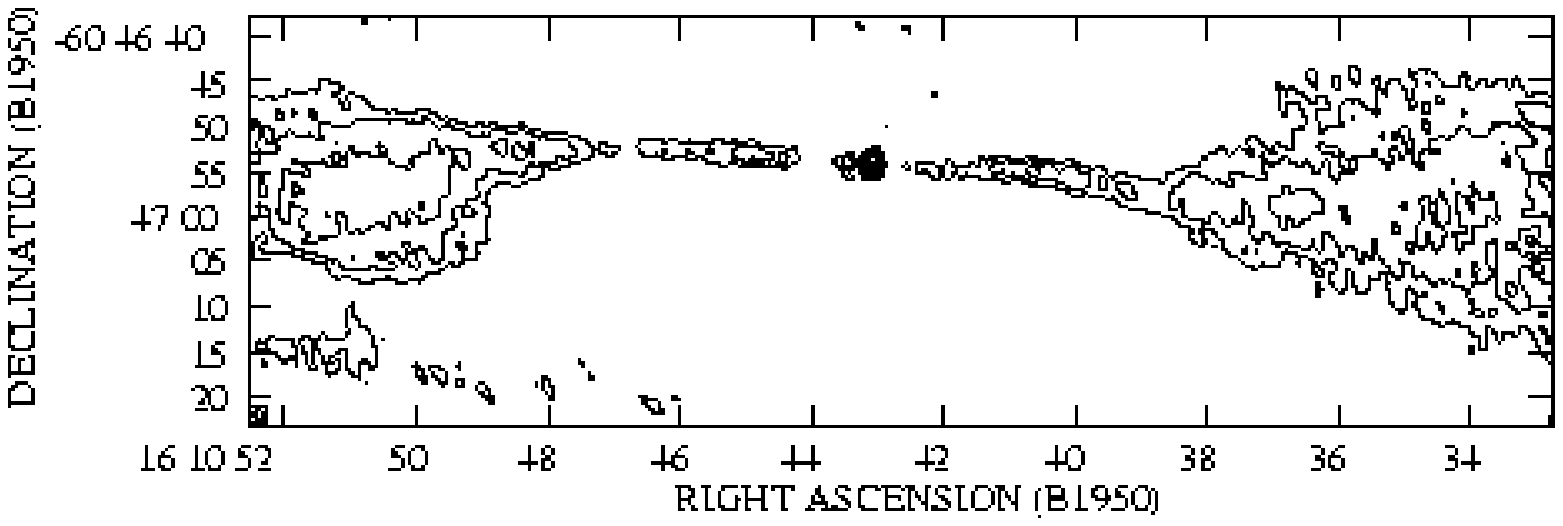}
\epsfxsize 13cm
\epsfbox{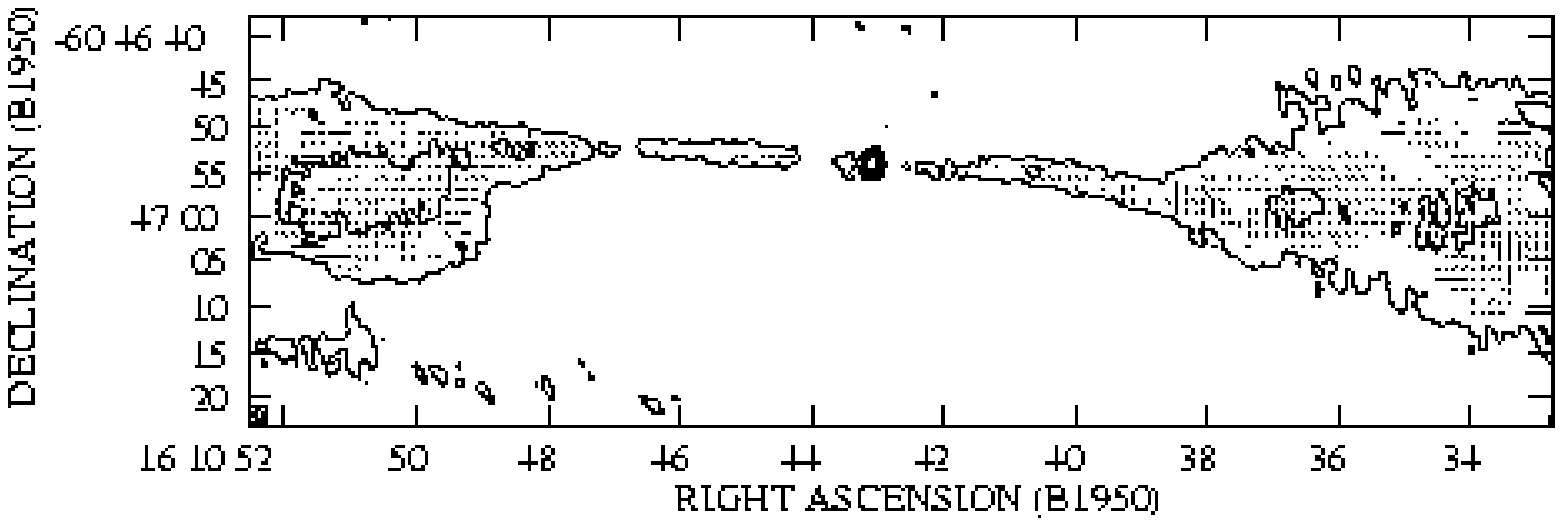}
\caption{5 and 8-GHz maps of \PKS . Top: C-band maps at $2.5 \times
1.8$ arcsec resolution: lowest contour is 2 mJy beam$^{-1}$ and
contours increase logarithmically by a factor 2 (upper panel) and 4
(lower panel). Bottom: X-band maps at $1.3 \times
1.0$ arcsec resolution: lowest contour is 0.8 mJy beam$^{-1}$,
contours as above.}
\label{1610}
\end{figure*}

\begin{figure}
\epsfxsize 8cm
\epsfbox{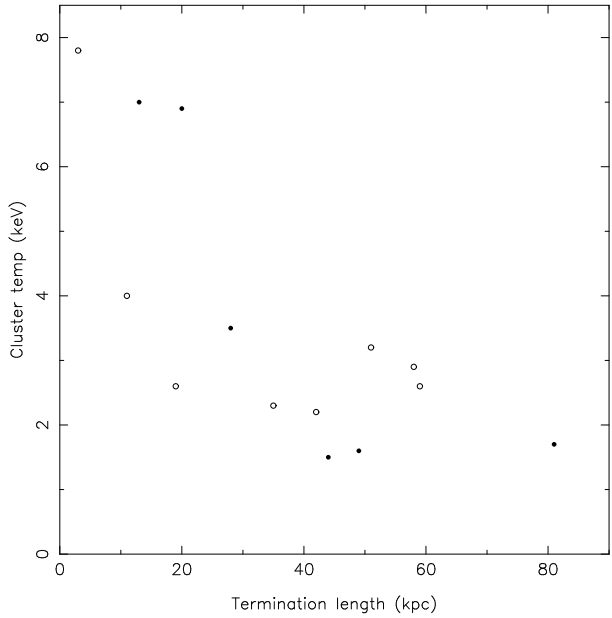}
\caption{Cluster temperature ($kT$, keV) against jet termination
  length $L$ (kpc), as defined in the text, for the WAT host clusters
  listed in Table \ref{ctemp}. Filled circles denote sources in our
  own sample, while open circles denote sources taken from the literature.}
\label{ktlen}
\end{figure}

\appendix
\renewcommand{\thefigure}{A\arabic{figure}}
\begin{figure*}
\epsfxsize 11cm
\epsfbox{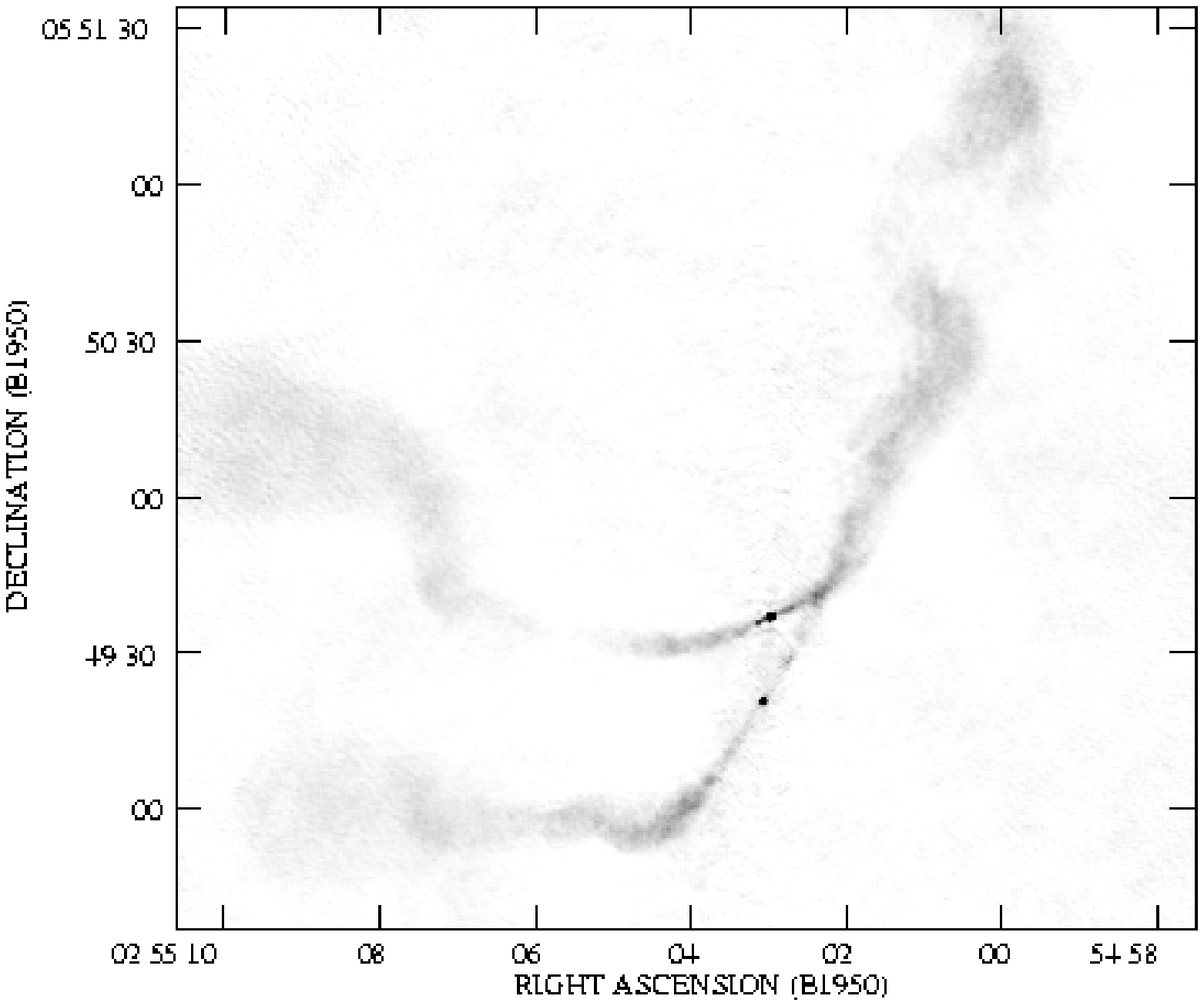}
\epsfxsize 11cm
\epsfbox{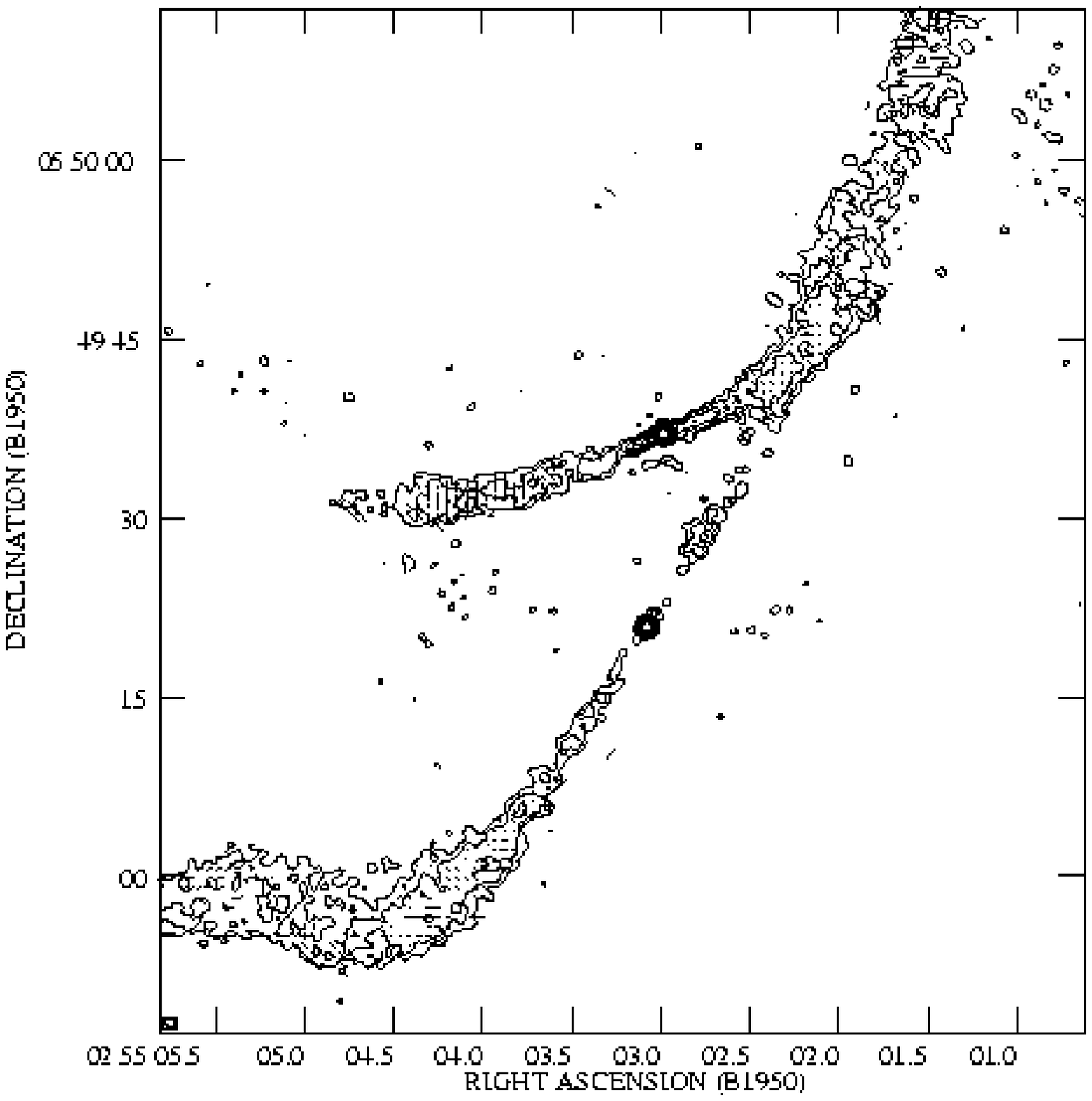}
\caption{$0.77 \times 0.73$-arcsec resolution map of the double
  twin-jet source 3C\,75. Top: the
  jets and plumes.
  Black is 1 mJy beam$^{-1}$. Bottom: the inner jets. As Fig.\
\ref{0647l}, but lowest contour level is 90 $\mu$Jy beam$^{-1}$.
Contours increase by a factor 2.}
\label{3c75}
\end{figure*}
\clearpage
\end{document}